\documentclass[a4paper,10pt]{article}
\usepackage[utf8]{inputenc}
\usepackage{amsthm}
\usepackage{amsfonts}
\usepackage{amssymb}	
\usepackage{amsmath}
\allowdisplaybreaks
\usepackage{mathtools}
\usepackage[english]{babel}
\usepackage{color}
\usepackage{slashed}
\usepackage{enumerate}
\usepackage{graphicx}
\usepackage{systeme}
\usepackage{dsfont}
\usepackage{relsize}
\usepackage[margin=0.90in]{geometry}
\usepackage{float}
\usepackage{tikz}
\usepackage[labelformat=simple]{subcaption}

\usepackage{graphicx}
\usepackage{bmpsize}
\usepackage{epstopdf}
\usepackage{bbm}
\usepackage{xcolor,etoolbox}
\usepackage{titling}
\usepackage{authblk}
\newcommand*\samethanks[1][\value{footnote}]{\footnotemark[#1]}
\usepackage{blindtext,graphicx}
\usepackage[absolute]{textpos}
\usepackage{physics}
\usepackage[hang,flushmargin]{footmisc}

\usepackage{xfrac}
\usepackage{nicefrac}
\usepackage{esvect}
\usepackage{cases}
\usepackage{empheq}
\usepackage{stmaryrd}
\usepackage{cancel}
\usepackage[linguistics]{forest}
\usepackage{url}
\usepackage[font=small]{caption}
\usepackage[giveninits=true,sortcites=true,date=year,maxbibnames=99,doi=false,isbn=false,url=false,eprint=false]{biblatex}
\renewbibmacro{in:}{}
\DeclareFieldFormat{pages}{#1}
\AtEveryBibitem{%
  \clearlist{language}%
}
\usepackage{csquotes}

\usepackage{setspace}
\makeatletter
\def\hlinewd#1{%
\noalign{\ifnum0=`}\fi\hrule \@height #1 \futurelet
\reserved@a\@xhline}
\makeatother
\theoremstyle{definition}

\makeatletter
\newcommand{\pushright}[1]{\ifmeasuring@#1\else\omit\hfill$\displaystyle#1$\fi\ignorespaces}
\newcommand{\pushleft}[1]{\ifmeasuring@#1\else\omit$\displaystyle#1$\hfill\fi\ignorespaces}
\makeatother
\title{Boundary and interface conditions in the relaxed micromorphic model: exploring finite-size metastructures for elastic wave control}
\author{
	Gianluca Rizzi\thanks{GEOMAS, INSA-Lyon, Universit\'e de Lyon, 20 avenue Albert Einstein,	69621, Villeurbanne cedex, France},
	\quad
	Marco Valerio d'Agostino\samethanks[1],
	\quad
	Patrizio Neff\thanks{Head of Chair for Nonlinear Analysis and Modelling, Fakultät für Mathematik, Universität Duisburg-Essen, \\ \indent Thea-Leymann-Straße 9, 45127 Essen, Germany},
	\quad and \quad
	Angela Madeo\samethanks[1]
	}

\thanksmarkseries{arabic}
\date{\today}
\begin{document}
\maketitle
\begin{abstract}
In this paper, we establish well-posed boundary and interface conditions for the relaxed micromorphic model that are able to unveil the scattering response of fully finite-size metamaterials' samples. The resulting relaxed micromorphic boundary value problem is implemented in finite element simulations describing the scattering of a square metamaterial's sample whose side counts 9 unit cells. The results are validated against a direct finite element simulation encoding all the details of the underlying metamaterial's microstructure. The relaxed micromorphic model can recover the scattering metamaterial's behavior for a wide range of frequencies and for all possible angles of incidence, thus showing that it is suitable to describe dynamic anisotropy.
Finally, thanks to the model's computational performances, we can design a metastructure combining metamaterials and classical materials in such a way that it acts as a protection device while providing energy focusing in specific collection points. These results open important perspectives for the short-term design of sustainable structures that can control elastic waves and recover energy.
\end{abstract}
\textbf{Keywords}: finite-size, metamaterials, metastructure, relaxed micromorphic model, wave channelling, wave focusing, energy harvesting.

\section{Introduction}
Mechanical metamaterials are architectured materials that may show unorthodox static and dynamic properties, due to their heterogeneous microstructures.
One of the main streams of research concerning mechanical metamaterials is focusing on how to create new micro-structures that give rise to exotic dynamic responses, so that  it is today possible to find researchers working on metamaterials exhibiting band-gaps \cite{celli_bandgap_2019,bilal_architected_2018,liu_locally_2000,wang_harnessing_2014}, cloaking \cite{buckmann_mechanical_2015,misseroni_cymatics_2016,zhou_elastic_2008,zhang_asymmetric_2020}, focusing \cite{cummer_controlling_2016,guenneau_acoustic_2007}, channeling \cite{kaina_slow_2017,tallarico_edge_2017}, negative refraction \cite{willis_negative_2016}, noise and vibration attenuation \cite{wen_effects_2011,mitchell_metaconcrete_2014,zhu_total-internal-reflection_2018}, super-resolution imaging \cite{zhu_negative_2014,kaina_negative_2015}, energy harvesting \cite{gonella_interplay_2009}, and many others.

Common numerical approaches to study the dynamical metamaterials' response include Bloch-Floquet analysis \cite{deymier_acoustic_2013,phani_wave_2006,alberdi_isogeometric_2018} and direct numerical simulations encoding all the details of the considered metamaterials' microstructures in a Finite Element environment \cite{krushynska_coupling_2017,krodel_wide_2015,liu_broadband_2018,celli_bandgap_2019,an_3d_2020}.
While Bloch-Floquet analysis is limited to the study of elastic wave propagation in unbounded domains, direct numerical simulations can precisely describe the mechanical metamaterials' behavior, also when considering finite-size samples.
However, these models may be so computationally expensive that they cannot be conveniently used to investigate metamaterials' response in different configurations combining metamaterials and classical materials.
These computational constraints are preventing us from designing large-scale metastructures that can have a true impact in real-world applications.

The awareness of these limitations triggered all the recent advances on dynamical homogenization methods \cite{chen_dispersive_2001,willis_exact_2009,craster_high-frequency_2010,willis_effective_2011,willis_construction_2012,boutin_large_2014,sridhar_general_2018}.
Such methods share the idea that a periodic infinite-size metamaterial can be replaced by a homogenized continuum, mimicking its response without accounting for all the microstructures' details.
This leads to important simplifications of metamaterials' modeling at the macroscopic scale, while still accounting for some influence of the underlying microstructure.

The stream of research on dynamical homogenization methods for metamaterials starts with \cite{willis_variational_1981,willis_exact_2009,willis_effective_2011} and is continued, among others, by \cite{nemat-nasser_overall_2011,srivastava_overall_2012,norris_analytical_2012,srivastava_limit_2014,nassar_willis_2015}.
While the homogenized elastic parameters obtained through this approach are capable of capturing the behavior of lower (acoustic and first optic) branches of the dispersion curves they cannot recover metamaterials' response at higher frequencies.

Other important attempts to generalize homogenization techniques to the dynamical framework relax the long wavelength assumption by making use of suitable asymptotic expansions \cite{nolde_high_2011,boutin_large_2014,bacigalupo_second-gradient_2014,hu_nonlocal_2017,hui_high_2014}.
Also in this case, the proposed homogenization frameworks can only recover metamaterials' response at relatively low frequencies (acoustic and first optic branches of the dispersion curves), or can locally get to higher branches of the dispersion curves when fixing a priori specific frequency values.

Computational homogenization theories based on scale separation hypotheses and extending the classical Hill-Mandel condition have been proposed e.g. in \cite{pham_transient_2013,sridhar_homogenization_2016,roca_computational_2018,sridhar_general_2018}.
The resulting frameworks can describe dispersive metamaterials behavior, but are still limited to low frequency and/or local resonance phenomena.

All this research effort has shown the importance of introducing homogenized models with effective elastic parameters to reduce otherwise prohibitive computational costs in dynamic simulations for metamaterials.
Homogenization methods give important indications about the dynamic response of unbounded metamaterials' specimens, but face the open problem of establishing well-posed boundary conditions at the macroscopic scale.
It is in fact very hard to set up homogenized boundary value problems that are i) directly derived from a homogenization procedure and ii) descriptive of realistic finite-size metamaterials' behavior.
The problems faced by classical homogenization methods for finite-size metamaterials' modeling have been very recently acknowledged by the cutting-edge research groups in dynamical homogenization \cite{srivastava_evanescent_2017,sridhar_homogenization_2016}.

An alternative to the homogenization approaches described so far is that of using so-called micromorphic models. These models were originally introduced by Mindlin \cite{mindlin_micro-structure_1964} and Eringen \cite{eringen_mechanics_1968} by extending the kinematics of classical Cauchy continua with extra degrees of freedom accounting for independent motions at the microscopic scale.
Thanks to this kinematic extension, micromorphic models can potentially describe dispersive metamaterials' behaviors in higher branches of the dispersion curves.

Some researchers recently tried to adapt existing up-scaling techniques so as to reach  micromorphic-like homogenized equations at the macroscopic scale.
Sridhar et al. \cite{sridhar_homogenization_2016} proposed an alternative ad-hoc up-scaling procedure, only valid for locally resonant metamaterials, leading to a homogenized equation which the authors recognize to be of the micromorphic type.
In a similar spirit, \cite{sridhar_frequency_2020} obtained a homogenized continuum with extended kinematics, classifying it as micromorphic, and proposed its use to study a simple 1D boundary value problem for a periodic metamaterial.
However, major concerns are always encountered when trying to establish well-posed homogenized boundary conditions.

Some of the authors of the present paper contributed to pioneer the so-called relaxed-micromorphic model \cite{neff_unifying_2014} and subsequently showed that it can be used to describe band-gap behaviors \cite{madeo_wave_2015,madeo_band_2015} that are recurrently observed in mechanical metamaterials (e.g., \cite{deymier_acoustic_2013,gantzounis_granular_2013,hussein_dynamics_2014,wang_harnessing_2014,cummer_controlling_2016}).
The stream of works that followed the creation of the relaxed micromorphic model \cite{madeo_wave_2015,madeo_band_2015,aivaliotis_microstructure-related_2019,neff_identification_2020,dagostino_effective_2020,aivaliotis_frequency-_2020,neff_relaxed_2015,madeo_reflection_2016} clearly indicates that a model that is not constrained to the classical homogenization mindset can be effectively engineered to catch complex metamaterials' responses, while keeping a reduced structure (free of unnecessary parameters).
Mathematical existence and uniqueness results have also been studied for this model \cite{neff_relaxed_2015}.
Subsequent works have focused on searching for an optimized structure of the relaxed-micromorphic constitutive laws to characterize realistic 1D metamaterials \cite{madeo_first_2016,madeo_relaxed_2018}.
Semi-analytical solutions for the frequency-dependent scattering of a relaxed-micromorphic half-plane where then provided \cite{aivaliotis_microstructure-related_2019}, thus acquiring deeper understanding on the fundamental problem of establishing well-posed boundary conditions in micromorphic media.
It was also shown (for the first time in these explicit terms) that the constitutive form of kinetic \cite{madeo_role_2017} and strain \cite{neff_identification_2020,dagostino_effective_2020} micromorphic energies is extremely important to describe essential metamaterials' features such as low- and high-frequency dispersion and dynamic anisotropy \cite{barbagallo_transparent_2017}.
It was then proven that the relaxed micromorphic model can describe the average behavior of certain infinite-size 2D metamaterials \cite{madeo_modeling_2018,barbagallo_relaxed_2019}.
Finally, some of the authors started exploring how boundary conditions should be introduced in micromorphic media to provide well-posed boundary value problems for 2D finite-size tetragonal metamaterials \cite{aivaliotis_microstructure-related_2019,aivaliotis_frequency-_2020,madeo_relaxed_2018}.

In the present paper, we focus on the major problem of establishing boundary and interface conditions that are i) intrinsically compatible with the relaxed micromorphic bulk PDEs and ii) able to describe ``realistic'' metamaterials'  dynamic responses when considering specimens of finite size.
Thanks to the computational performances of the relaxed micromorphic model, coupled with the introduction of such well-posed boundary and interface conditions, we are able to design and optimize a complex metastructure for protection applications that also allows energy focusing for eventual energy recovery.
This metastructure is made of  metamaterials and classical (homogeneous) materials bricks that are combined together in such a way that the structure acts as a protection device, while being able to focus energy in specific paths for eventual subsequent re-use.

The paper is now structured as follows. In Section 1 we present an introduction that makes a point about the current state of the art about metamaterials' modelling and about the relaxed micromorphic modelling of metamaterials.

Section 2 recalls the bulk equations for both Cauchy and relaxed micromorphic continua and then focuses on the establishment of well-posed boundary and interface duality conditions that can describe ``realistic'' metamaterials' boundaries, as well as interfaces between different metamaterials and metamaterials/homogeneous materials.

In section 3 we present numerical simulations showing the performances of the relaxed micromorphic model to describe the scattering of finite-size metamaterial's sample.
The obtained results clearly indicate the capability of the relaxed micromorphic model of correctly describing metamaterials's dynamic anisotropy.

In section 4 we exploit the computational performances of the relaxed micromorphic model to explore and design a metastructure that acts as a protection tool, while being able to focus energy in specific paths for eventual subsequent reuse.

In section 5, we draw our conclusions and perspective for subsequent works.
\section{Relaxed micromorphic modelling of finite-size metamaterials}
\label{sec:const_law}
In this section, we briefly recall the weak and strong form of the governing equations of both classical Cauchy and relaxed micromorphic continua.
We then put a particular focus on the establishment of well-posed boundary conditions that are indispensable for the realistic modeling of the free surface of finite-size metamaterial's specimens, as well as of the interfaces between different metamaterials and linear-elastic classical Cauchy materials.

The Lagrangian $\mathcal{L}$ for the classical Cauchy model is
\begin{align}
\mathcal{L} \left(\nabla u,u_{,t}\right) =
\dfrac{1}{2}\rho \, \langle u_{,t},u_{,t} \rangle
-
\dfrac{1}{2} \langle \mathbb{C} \, \mbox{sym}\nabla u, \mbox{sym}\nabla u \rangle
\, ,
\label{eq:class_energy}
\end{align}
where $u$ is the displacement field, $(\cdot)_{,t}$ is intended as a derivative with respect to time, $\rho$ is the apparent mass density, and $\mathbb{C}$ is the classical 4th order elasticity tensor.
The Lagrangian $\mathcal{L}_{\text{m}}$ for the relaxed micromorphic model is \cite{aivaliotis_frequency-_2020}
\begin{align}
\mathcal{L}_{\text{m}} \left(u_{,t},\nabla u_{,t}, P_{,t}, \nabla u, P, \mbox{Curl}\, P\right) =
& 
\, 
\dfrac{1}{2}\rho \, \langle u_{,t},u_{,t} \rangle + 
\dfrac{1}{2} \langle \mathbb{J}_{\text{micro}} \, \mbox{sym} \, P_{,t}, \mbox{sym} \, P_{,t} \, \rangle 
+ \dfrac{1}{2} \langle \mathbb{J}_{\text{c}} \, \mbox{skew} \, P_{,t}, \mbox{skew} \, P_{,t} \rangle
\notag
\\*
&
+ \dfrac{1}{2} \langle \mathbb{T}_{\text{e}} \, \mbox{sym}\nabla u_{,t}, \mbox{sym}\nabla u_{,t} \rangle
+ \dfrac{1}{2} \langle \mathbb{T}_{\text{c}} \, \mbox{skew}\nabla u_{,t}, \mbox{skew}\nabla u_{,t} \rangle
\label{eq:relax_energy}
\\*
&
-\dfrac{1}{2} \langle \mathbb{C}_{\text{e}} \mbox{sym}\left(\nabla u - P \right), \mbox{sym}\left(\nabla u - P \right) \rangle
- \dfrac{1}{2} \langle \mathbb{C}_{\text{micro}} \mbox{sym} P,\mbox{sym} P \rangle
\notag
\\*
&
- \dfrac{1}{2} \langle \mathbb{C}_{\text{c}} \mbox{skew}\left(\nabla u - P \right), \mbox{skew}\left(\nabla u - P \right) \rangle
- \dfrac{1}{2} \langle \mathbb{L} \, \mbox{Curl}\, P, \mbox{Curl}\, P \rangle
\notag
\end{align}
where $u \in \mathbb{R}^{3}$ is the macroscopic displacement field, $P \in \mathbb{R}^{3\times 3}$ is the non-symmetric micro-distortion tensor, $\rho$ is the macroscopic apparent density, $\mathbb{J}_{\text{micro}}$, $\mathbb{J}_{\text{c}}$, $\mathbb{T}_{\text{e}}$, $\mathbb{T}_{\text{c}}$, are 4th order micro-inertia tensors, and $\mathbb{C}_{\text{e}}$, $\mathbb{C}_{\text{micro}}$, $\mathbb{C}_{\text{c}}$, $\mathbb{L}$ are 4th order elasticity tensors.
In the numerical applications considered in this work, all the possible characteristic lengths contained in the tensor $\mathbb{L}$ will be set equal to zero since their effect can, in a first instance, be neglected when focusing on the dynamic regime.
The action functional $\mathcal{A}$ of the considered continuum can be defined based on the Lagrangian function as
\begin{align}
    \mathcal{A}=\int\limits_{\Omega \times \left[0,T\right]} \mathcal{L} \,\, dx \, dt \, ,
    \quad\quad\quad
    \mbox{and}
    \quad\quad\quad
    \mathcal{A}_{\text{m}}=\int\limits_{\Omega \times \left[0,T\right]} \mathcal{L}_{\text{m}} \,\, dx \, dt \, ,
    \label{eq:action_func}
\end{align}
for classical Cauchy and relaxed micromorphic media, respectively.
In eq.(\ref{eq:action_func}), $\Omega$ stands for the volume of the considered continuum in its reference configuration and $[0,T]$ is a time interval during which the deformation of the continuum is observed.
In the case of conservative internal actions, the virtual work of internal actions $\mathcal{W}^{\text{int}}$ can be defined based on the first variation of the action functional.
In formulas, we have:
\begin{align}
    \int_{0}^{T}\, \mathcal{W}^{\text{int}} \, dt
    =
    \delta \mathcal{A} 
    \, ,
    \quad\quad\quad
    \mbox{and}
    \quad\quad\quad
    \int_{0}^{T} \,  \mathcal{W}_{\text{m}}^{\text{int}} \, dt 
    =
    \delta \mathcal{A}_{\text{m}}
    \, ,
    \label{eq:first_var_A}
\end{align}
for classical Cauchy and relaxed micromorphic media, respectively.
In eq.(\ref{eq:first_var_A}) the variation operator $\delta$ indicates that the variation must be taken with respect to the unknown kinematics fields ($u$ for Cauchy media and ($u,P$) for the relaxed micromorphic model).
Following classical variational calculus, the strong form of the bulk equations of motion for Cauchy and the relaxed micromorphic model can be obtained via a least action principle stating that, in absence of external body loads, the first variation of the action functional must be vanishing.
The application of a least-action principle to isotropic Cauchy and relaxed micromorphic models respectively gives \cite{rizzi_exploring_2021,rizzi2020towards,aivaliotis_frequency-_2020,neff_unifying_2014}
\begin{align}
    \rho \, u_{,tt} = \mbox{Div}\, \sigma \, ,
    \qquad\qquad\qquad
    \sigma
    =
    2\mu \, \mbox{sym}\nabla u + \lambda \, \mbox{tr}\left(\mbox{sym} \nabla u\right) \, \boldsymbol{\mathbbm{1}}
\label{eq:equiCau}
\end{align}
and
\begin{equation}
\rho \, u_{,tt} - \mbox{Div}\left(\widehat{\sigma}_{,tt}\right) = \mbox{Div}\left(\widetilde{\sigma}\right) \, ,
\qquad
\left( \mathbb{J}_{\text{micro}} + \mathbb{J}_{\text{c}} \right) \, P_{,tt} = \widetilde{\sigma} - s -\text{Curl} \, m \, .
\label{eq:equiMic}
\end{equation}
where
\begin{gather}
    \widetilde{\sigma} \coloneqq \mathbb{C}_{\text{e}}~\mbox{sym}\left(\nabla u -  \, P \right) + \mathbb{C}_{\text{c}}~\mbox{skew}\left(\nabla u -  \, P \right) \, ,
    \qquad
    \widehat{\sigma} \coloneqq \mathbb{T}_{\text{e}}~\mbox{sym} \nabla u + \mathbb{T}_{\text{c}}~\mbox{skew} \nabla u \, , 
    \label{eq:equiSigAll}
    \\*[2mm]
    s \coloneqq \mathbb{C}_{\text{micro}}\, \text{sym} \, P \, ,
    \qquad
    m \coloneqq \mathbb{L} \, \text{Curl} \, P \, .
    \notag
\end{gather}
\subsection{Relaxed micromorphic boundary and interface conditions that are representative of realistic metamaterial's interfaces}
The fact of establishing well-posed boundary conditions for the relaxed micromorphic model that are representative of realistic physical situations is of major importance for its sound application in view of the conception of complex metastructures that can control elastic waves and recover energy.

To obtain the conditions that can be established at the boundary $\partial\Omega$ of a relaxed micromorphic continuum, the work of external actions $\mathcal{W}^{\text{ext}}_{\text{m}}$ must be introduced as
\begin{align}
    \int_{0}^{T} \, \mathcal{W}^{\text{ext}}_{\text{m}} \, dt
    =
    \int\limits_{\partial\Omega \times \left[0,T\right]} \langle f^{\text{ext}} , \delta u \rangle \, dx \, dt
    +
    \int\limits_{\partial\Omega \times \left[0,T\right]} \langle \Phi^{\text{ext}} , \delta P \rangle \, dx \, dt \, ,
\end{align}
where $f^{\text{ext}}$ and $\Phi^{\text{ext}}$ are the external forces and double forces acting through the boundary $\partial \Omega$ of the considered domain and $\delta u$ and $\delta P$ are the test functions associated to the micromorphic kinematic fields.
\\
\indent
The duality conditions to be imposed at the boundary $\partial \Omega$ of the relaxed micromorphic model can be unveiled via the principle of virtual works which states that
\begin{align}
    \mathcal{W}^{\text{int}}_{\text{m}} = \mathcal{W}^{\text{ext}}_{\text{m}} \, .
\end{align}
In this way, the weak conditions that must be verified on the boundary $\partial \Omega$ read
\begin{align}
    \langle t_{\text{m}}, \delta u \rangle = \langle f^{\text{ext}}, \delta u \rangle \, ,
    \qquad\qquad\qquad
    \langle \tau, \delta P \rangle = \langle \Phi^{\text{ext}}, \delta P \rangle
    \label{eq:trac_ext_int}
\end{align}
where $t_{\text{m}}$ and $\tau$ are the generalized traction and double traction defined as
\begin{align}
t_{\text{m}} = \left(\widetilde{\sigma} + \widehat{\sigma}_{,tt} \right) \nu \, ,
\qquad\qquad\qquad
\tau = m \times \nu \, ,
\label{eq:sigMic}
\end{align}
with $\nu$ the normal to the considered boundary $\partial \Omega$, and where the cross product $\times$ is intended row-wise.
It can be shown that, in absence of curvature terms ($\mathbb{L}=0$), the duality condition condition involving $\Phi^{\text{ext}}$ and $\delta P$ must not be assigned on the boundary.

There are different possible ways to act on the boundary of the relaxed micromorphic continuum (e.g. assign force or displacement), but in all cases such conditions must verify eq.(\ref{eq:trac_ext_int}).
It is clear that the duality boundary conditions eq.(\ref{eq:trac_ext_int}) are modified if, instead of having a free metamaterial's boundary, we consider an interface between two metamaterials or between a metamaterial and a Cauchy material.
It can be shown that in this case, the interface duality conditions read
\!\!\!
\footnote{
Here and in the sequel, we suppose that no additional external surface forces are present when considering ``internal'' interfaces between two continua.
}
\begin{align}
    \langle t^+, \delta u^+\rangle
    =
    \langle t^-, \delta u^-\rangle \, ,
    \qquad\qquad
    \langle \tau^+, \delta P^+\rangle
    =
    \langle \tau^-, \delta P^-\rangle \, ,
    \label{eq:trac_ext_int_2}
\end{align}
where $t^+$ $t^-$, $\tau^+$, and $\tau^-$ are the tractions and the double tractions at the considered interface, as obtained taking the limit from the ``+'' and ``-'' side, respectively.
Depending on the fact that the ``+'' and the ``-'' regions of the domain are occupied by a relaxed micromorphic continuum or by a classical Cauchy continuum, the traction will be ``generalized'' or ``classical'' traction, respectively, while the double traction will be either given by eq.(\ref{eq:sigMic})$_2$ or vanishing.

For the sake of completeness, we recall here the definition of the classical Cauchy traction for the isotropic case
\begin{align}
    t=\sigma \, \nu \, ,
    \qquad
    \sigma = 2\mu \, \mbox{sym}\nabla u + \lambda \, \mbox{tr}\left(\mbox{sym} \nabla u\right) \, \boldsymbol{\mathbbm{1}}
    \, .
    \label{eq:sigClass}
\end{align}

In the reminder of this paper, we will show how the boundary and/or interface conditions that have to be used in the relaxed micromorphic framework to describe ``realistic'' metastructures must always be chosen to satisfy the duality conditions (\ref{eq:trac_ext_int}) and/or (\ref{eq:trac_ext_int_2}).

\section{Finite element implementation of a relaxed micromorphic boundary value problem for finite-size metamaterials}
\label{sec:finite_elem}
\begin{figure}[H]
	\centering
	\includegraphics[height=6cm]{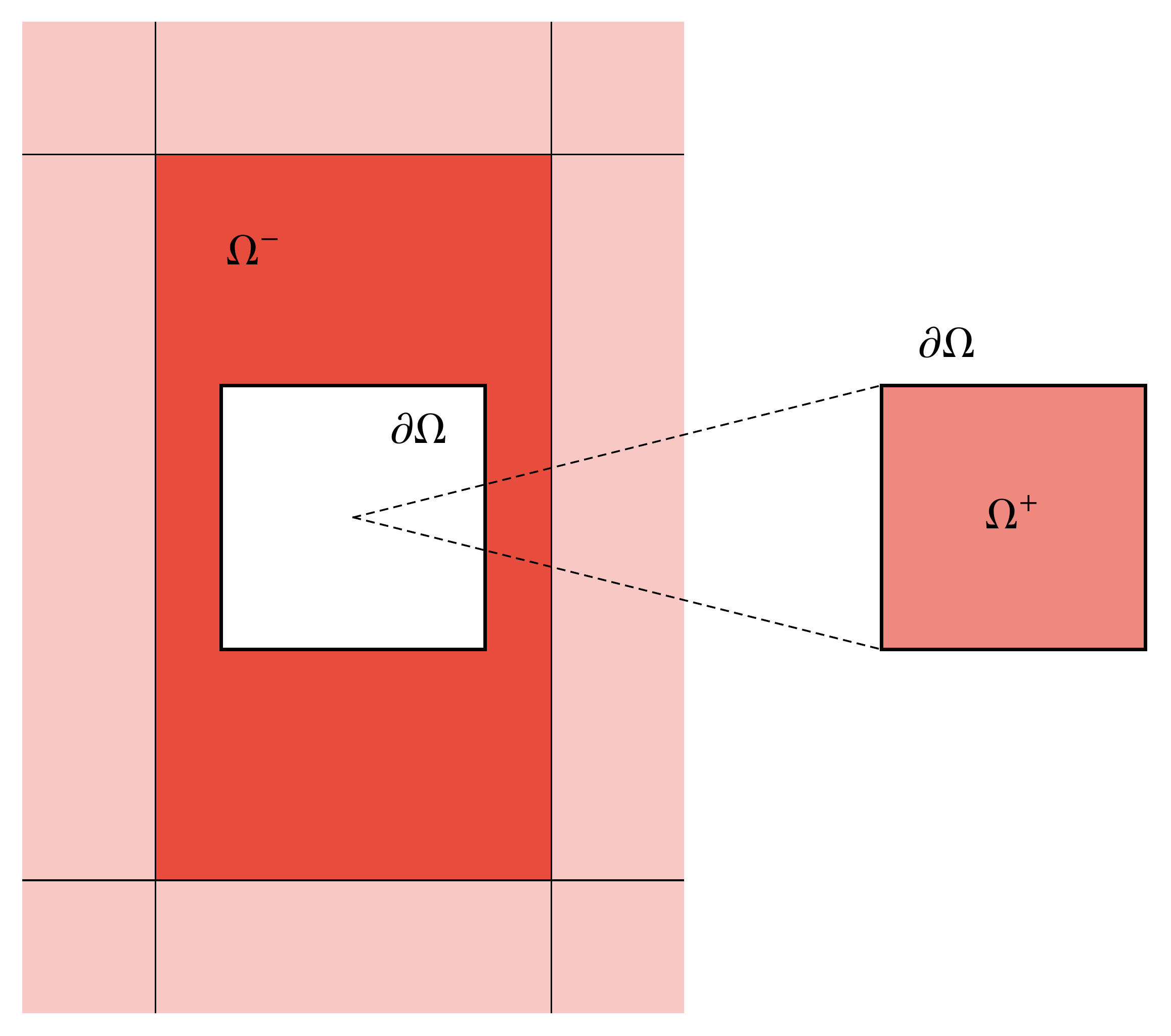}
	\caption{Schematic representation of the finite-size relaxed micromorphic continuum ($\Omega^{+}$) embedded in an infinite Cauchy continuum ($\Omega^{-}$).}
	\label{fig:jump_conditions}
\end{figure}

All the finite element analyses presented in this paper have been performed in the \textit{time harmonic domain} (plane incident wave) and the following interface conditions on the displacement and on the traction have been enforced at the interface $\partial \Omega$ (see Fig.~\ref{fig:jump_conditions}):
\!\!\!\!
\footnote{
Since we assume $\mathbb{L}=0$, the conditions on the micro-distortion $P$ and on the double stress $\tau$ must not be assigned here.
}
\begin{align}
\left\{
\begin{array}{rrrrrrrrrr}
u^{-} 
& = 
& u^{+} \, ,
\\*[2mm]
 t^{-} 
& = 
& t^{+} \, ,
\end{array}
\right.
\quad \mbox{on} 
\quad \partial \Omega \, ,
\label{eq:trac_1}
\end{align}
where $u^{-}$, $t^{-}$, $u^{+}$, and $t^{+}$ are the displacements and the tractions on the interface $\partial \Omega$, computed as the limit from the ``-'' and the ``+'' side, respectively.
It is clear that these strong-form boundary conditions, also enforce the validity of their weak counterparts (\ref{eq:trac_ext_int_2})$_1$, so that the considered differential problem is well-posed.

We stress once again that, since the region $\Omega^{-}$ is occupied by a Cauchy continuum and the region $\Omega^{+}$ is occupied by a relaxed micromorphic continuum, the tractions $t^{-}$ and $t^{+}$ are respectively given by 
\begin{align}
t^{-} = \sigma \, \nu \, ,
\qquad\qquad\qquad
t^{+} = \left(\widetilde{\sigma} + \widehat{\sigma}_{,tt} \right) \nu \, .
\end{align}
Taking advantage of the superposition principle and assuming that an incident wave is travelling in the domain $\Omega^{-}$, it is possible to decompose the external displacement $u^-$ and traction $t^-$ into an incident and scattered component, while the internal displacement $u^+$ only consists of a scattered component.
We can thus rewrite the interface conditions (\ref{eq:trac_1}) as
\begin{align}
\left\{
\begin{array}{ccccccccccccc}
 u^{-} 
& = 
& u^{+}
& \Longleftrightarrow 
& u^{-}_{\footnotesize \mbox{scat}} + u_{\footnotesize \mbox{inc}}
& = 
& u^{+}_{\footnotesize \mbox{scat}} \, ,
\\*[2mm]
 t^{-} 
& = 
& t^{+}
& \Longleftrightarrow 
& t^{-}_{\footnotesize \mbox{scat}} + t_{\footnotesize \mbox{inc}}
& = 
& t^{+}_{\footnotesize \mbox{scat}} \, ,
\end{array}
\right.
\quad \mbox{on} 
\quad \partial \Omega \, .
\label{eq:jump_cond}
\end{align}

In this way it is possible to assign the ``incident wave load'' at the interface by setting the jump between the internal and the external scattered field to be equal to the incident wave contribution.
This allows us to write 
\begin{align}
\begin{array}{ccccccccc}
u^{-}_{\footnotesize \mbox{scat}}  
& = 
& u^{+}_{\footnotesize \mbox{scat}} - u_{\footnotesize \mbox{inc}}, 
& \quad \mbox{on} 
& \quad \partial \Omega \, ,
\\*[3mm]
t^{-}_{\footnotesize \mbox{scat}}  
& = 
& t^{+}_{\footnotesize \mbox{scat}} - t_{\footnotesize \mbox{inc}}, 
& \quad \mbox{on} 
& \quad \partial \Omega \, .
\end{array}
\label{eq:jump_cond_2}
\end{align}
Therefore, the final unknowns determined by the solution of the finite-element analysis are the internal (in $\Omega^{+}$) and external (in $\Omega^{-}$) scattered displacement fields.

\subsection{Relaxed micromorphic model validation on the anisotropic scattering of a finite-size metamaterial's sample}
\begin{figure}[H]
\begin{minipage}{0.45\textwidth}
\centering
  \includegraphics[width=0.5\textwidth]{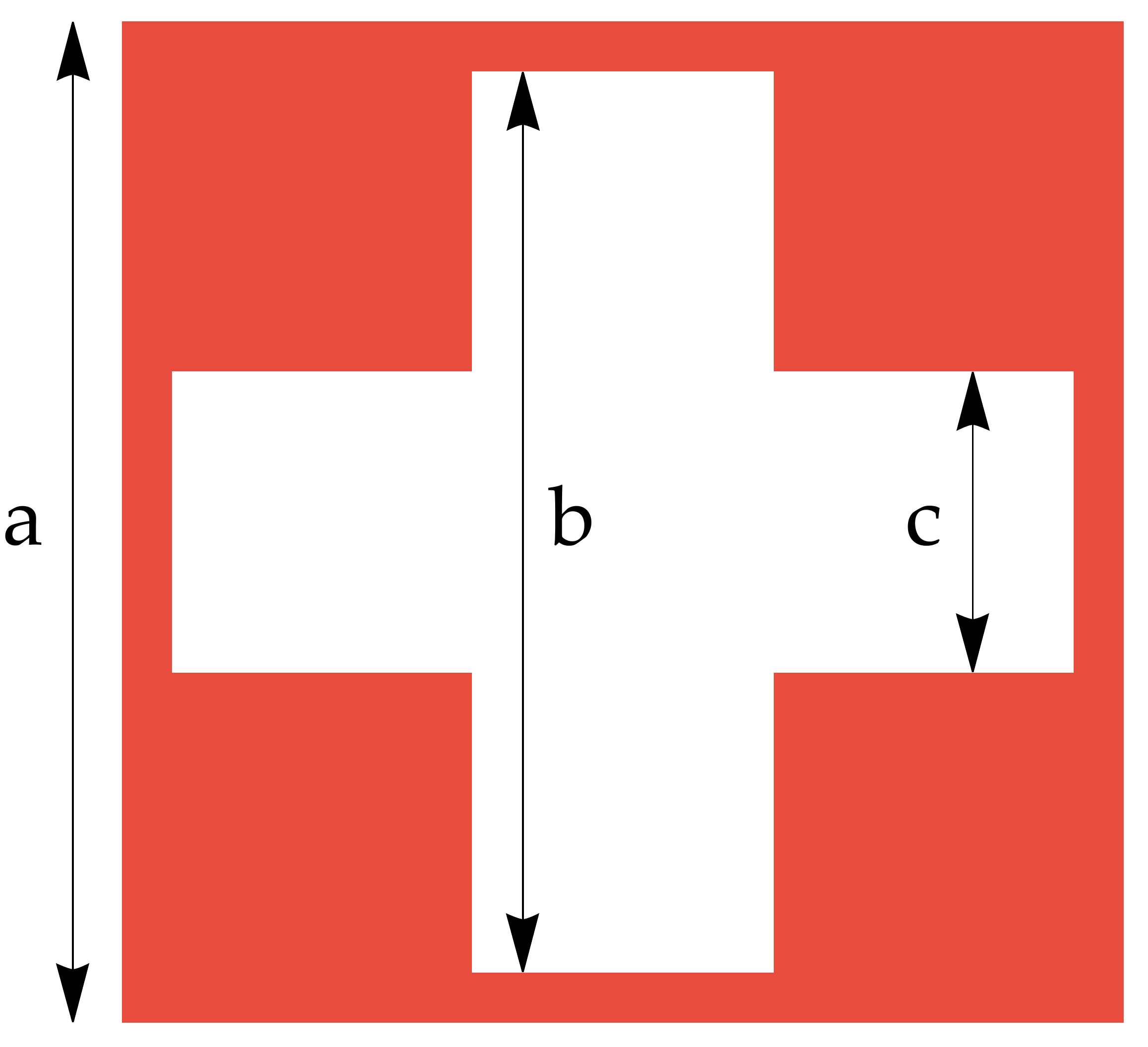}
    \caption{Unit cell of the investigated metamaterial. The matrix is made-up of aluminum (red).}
    \label{fig:unit_cell}
\end{minipage}
\hfill
\begin{minipage}{0.45\textwidth}
\centering
\captionof{table}{Unit cell's dimensions and material properties of aluminum.}
\begin{tabular}{rl||rl}
\hline
\hline
 a &  1.0 mm & $\lambda_{\text{Al}}$ & 51.1 [GPa]\\
 \hline
 b &  0.9 mm & $\mu_{\text{Al}}$ & 26.3 [GPa]\\
 \hline
 c &  0.3 mm & $\rho_{\text{Al}}$ & 2700 [kg/m$^3$]\\
 \hline
 \hline
\end{tabular}
\label{tab:unit_cell}
\end{minipage}
\end{figure}

In this section, we consider a square metamaterial's specimen whose side is constituted by 9 unit cells of the type depicted in Fig.~\ref{fig:unit_cell} and whose geometric and elastic properties are given in Table~\ref{tab:unit_cell}.

\begin{figure}[H]
	\centering
	\includegraphics[height=6cm]{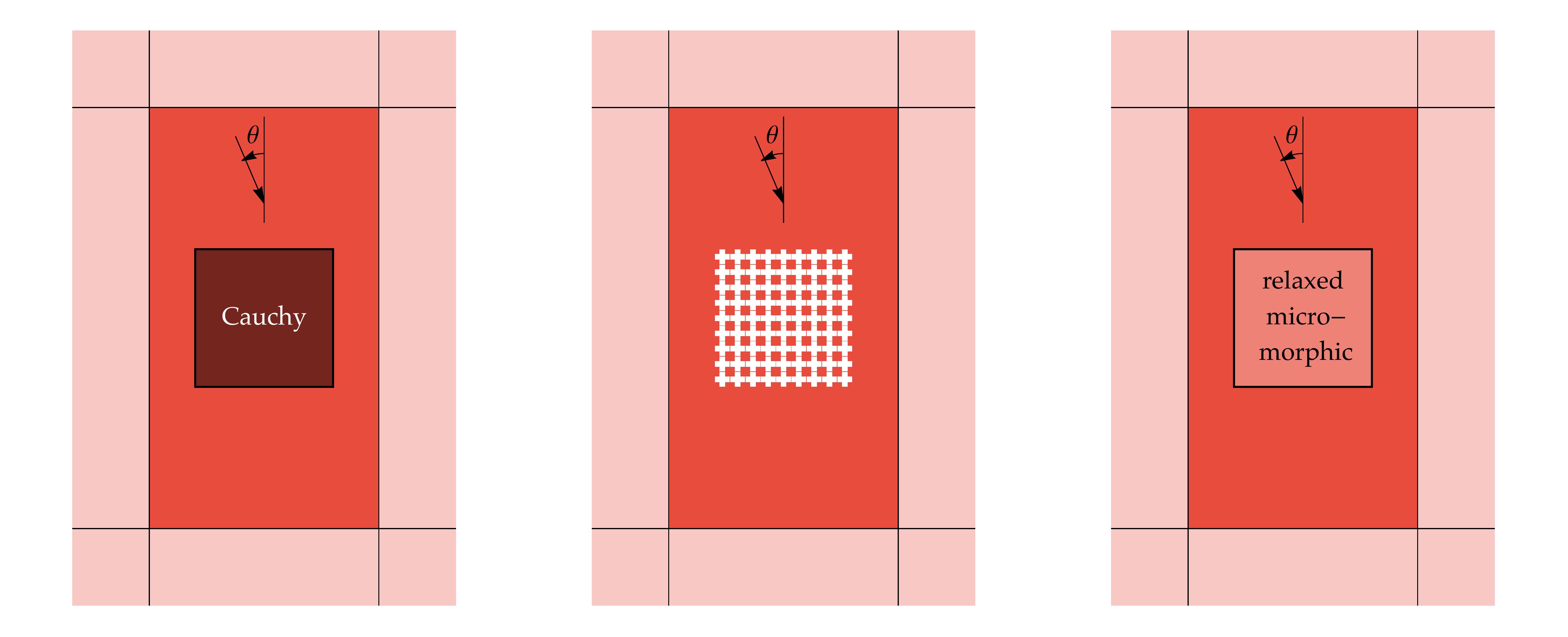}
	\caption{Schematic representation of the geometries implemented in the finite-element simulation for:
	(\textit{left}) the equivalent Cauchy sample;
	(\textit{centre}) a metamaterial's sample;
	(\textit{right}) a relaxed micromorphic sample.
	The incident wave comes from the external Cauchy medium and can be sent with an arbitrary angle $\theta$.}
\label{fig:geometry}
\end{figure}

This metamaterial's sample is embedded in a Cauchy continuum of the same type (elastic properties given in Table~\ref{tab:unit_cell}).
This means that the considered metastructure can be realized by drilling holes of the type presented in Fig.~\ref{fig:unit_cell} inside a homogeneous aluminum plate which is much larger than the internal metamaterial's sample (Fig.~\ref{fig:geometry}(\textit{center})).
The external Cauchy continuum is modelled as an infinite medium by the mean of using a perfectly matched layer (PML).
All the analyses are carried out under the plane-strain hypothesis.
Since there is no discontinuity of the material properties at the boundary between the external and internal material, no specific interface conditions must be imposed when implementing the detailed micro-structured finite-element simulation.
The scattered profile will be produced by the fact that the incident wave encounters the traction-free surface of the holes.
To validate the relaxed micromorphic model for the description of the scattering of this finite-size metamaterial's sample, we also implement the corresponding finite-element simulation in which the internal domain is modelled by means of the eq.(\ref{eq:relax_energy}) and the interface conditions (\ref{eq:jump_cond_2}) (Fig.~\ref{fig:geometry}(\textit{right})).
The plane-strain hypothesis is accounted for by setting $u_3=0$ and $P_{13}=P_{23}=P_{31}=P_{32}=P_{33}=0$.
The relaxed micromorphic parameters for the metamaterial of Fig.~\ref{fig:unit_cell} are defined in eq.(\ref{eq:micro_ine_1}), they have been calibrated in \cite{aivaliotis_frequency-_2020} on an infinite metamaterial, and their values are summarized in Table~\ref{tab:parameters_RM}.
\begin{equation}
\begin{array}{rlrl}
	\mathbb{J}_{\text{micro}} &=\rho 
	\begin{pmatrix}
	\eta_{3} + 2\eta_{1} & \eta_{3}              & \dots 			& \bullet \\ 
	\eta_{3}            & \eta_{3} + 2\eta_{1} & \dots 			& \bullet \\ 
	\vdots               & \vdots                 & \ddots 			& \bullet \\ 
	\bullet              & \bullet                & \bullet 		& \eta^{*}_{1} \\ 
	\end{pmatrix} \, ,
	\quad
	&\mathbb{J}_{\text{c}} &= \rho
	\begin{pmatrix}
	\bullet & 			& \bullet\\ 
	& \ddots 	& \vdots\\ 
	\bullet & \dots & 4\eta_{2}
	\end{pmatrix} \, ,
	\\[1.5cm]
	\mathbb{T}_{\text{e}} &= \rho
	\begin{pmatrix}
	\overline{\eta}_{3} + 2\overline{\eta}_{1}	& \overline{\eta}_{3}        			   	& \dots		& \bullet\\ 
	\overline{\eta}_{3}         				    & \overline{\eta}_{3} + 2\overline{\eta}_{1} 	& \dots		& \bullet\\ 
	\vdots                    					& \vdots                 				   	& \ddots 	&		 \\ 
	\bullet                   					& \bullet									& 	 		& \overline{\eta}^{*}_{1}
	\end{pmatrix} \, ,
	\quad
	&\mathbb{T}_{\text{c}} &= \rho
	\begin{pmatrix}
	\bullet & 			& \bullet\\ 
	& \ddots 	& \vdots\\ 
	\bullet & \dots & 4\overline{\eta}_{2}
	\end{pmatrix} \, ,
    \\[1.5cm]
    \mathbb{C}_{\text{e}} &= 
    \begin{pmatrix}
    \lambda_{\text{e}} + 2\mu_{\text{e}}	& \lambda_{\text{e}}				& \dots		& \bullet\\ 
    \lambda_{\text{e}}				& \lambda_{\text{e}} + 2\mu_{\text{e}}	& \dots		& \bullet\\ 
    \vdots					& \vdots					& \ddots	& 		 \\ 
    \bullet					& \bullet					& 			& \mu_{\text{e}}^{*}\\ 
    \end{pmatrix} \, ,
    \quad
    &\mathbb{C}_{\text{c}} &= 
    \begin{pmatrix}
    \bullet & 			& \bullet\\ 
    		& \ddots 	& \vdots\\ 
    \bullet & \dots		& 4\mu_{\text{c}}
    \end{pmatrix} \, ,
    \\[1.5cm]
    \mathbb{C}_{\text{micro}} &= 
    \begin{pmatrix}
    \lambda_{\text{micro}} + 2\mu_{\text{micro}}	& \lambda_{\text{micro}}				& \dots		& \bullet\\ 
    \lambda_{\text{micro}}				& \lambda_{\text{micro}} + 2\mu_{\text{micro}}	& \dots		& \bullet\\ 
    \vdots					& \vdots					& \ddots	&  \\ 
    \bullet					& \bullet					& 			& \mu_{\text{micro}}^{*}\\ 
    \end{pmatrix} \, ,
    \end{array}
\label{eq:micro_ine_1}
\end{equation}
In eq.(\ref{eq:micro_ine_1}) we report the elasticity tensors expressed in Voigt notation, where only the parameters involved under the plane-strain hypothesis are presented.

\begin{table}[H]
	\renewcommand{\arraystretch}{1.5}
	\centering
	\begin{subtable}[t]{.70\textwidth}
		\centering
		\begin{tabular}{ccccccc} 
			$\lambda_{\text{e}}$ [GPa] & $\mu_{\text{e}}$ [GPa] & $\mu^{*}_{\text{e}}$ [GPa]  & $\mu_{\text{c}}$ [GPa]  \\
			\hline\hline
			$2.33$ & $10.92$  & $0.67$ & $2.28\times10^{-3}$ \\
			\hlinewd{2pt}
			$\lambda_{\text{micro}}$ [GPa] & $\mu_{\text{micro}}$ [GPa] & $\mu^{*}_{\text{micro}}$ [GPa] & $\rho$ [kg/m$^3$]\\
			\hline\hline
			$5.27$  & $12.8$  & $8.33$ & $1485$ \\ 
			\hlinewd{2pt}
			$\eta_{1}$ [kg/m] & $\eta_{2}$ [kg/m] & $\eta_{3}$ [kg/m] & $\eta^{*}_{1}$ [kg/m]\\
			\hline\hline
			$8.6\times10^{-5}$ & $10^{-7}$ & $-2.2\times10^{-5}$ & $3.3\times10^{-5}$ \\ 
			\hlinewd{2pt}
			$\overline{\eta}_{1}$ [kg/m] & $\overline{\eta}_{2}$ [kg/m] & $\overline{\eta}_{3}$ [kg/m] & $\overline{\eta}^{*}_{1}$ [kg/m]\\
			\hline\hline
			$5.6\times10^{-5}$  & $7.3\times10^{-4}$ & $1.7\times10^{-4}$ & $9\times10^{-7}$ \\ 
		\end{tabular}
	\end{subtable}
	\hfill
	\centering
	\begin{subtable}[t]{.20\textwidth}
		\centering
		\begin{tabular}{ccc} 
			$\lambda_{\text{macro}}$ [GPa]\\
			\hline\hline
			$1.74$\\
			\hlinewd{2pt}
			$\mu_{\text{macro}}$ [GPa]\\
			\hline\hline
			$5.89$\\
			\hlinewd{2pt}
			$\mu^{*}_{\text{macro}}$ [GPa]\\
			\hline\hline
			$0.62$
		\end{tabular}
		\vspace{1.1cm}
	\end{subtable}
	\caption{
		In the (\textit{left}) table we report the values of the relaxed micromorphic static and dynamic parameters for the metamaterial in Fig~\ref{fig:unit_cell} determined via the fitting procedure introduced in \cite{dagostino_effective_2020,aivaliotis_frequency-_2020}. The apparent density $\rho$ is computed based on the aluminum  microstructure of Fig.~\ref{fig:unit_cell}.
		In the (\textit{right}) table are reported the values of the equivalent Cauchy continuum elastic coefficients corresponding to the long-wave limit of the relaxed tetragonal micromorphic model as obtained with the procedure explained in \cite{neff_identification_2020}.
	}
	\label{tab:parameters_RM}
\end{table}

The calibration procedure presented in \cite{aivaliotis_frequency-_2020} is performed by calibrating the relaxed micromorphic parameters on the two principal directions ($\theta=0$\textdegree  and $\theta=45$\textdegree).
However, due to the fact that the micro- and macro-tetragonal symmetry is taken into account through eq.(\ref{tab:parameters_RM}) of the relaxed micromorphic elasticity tensors, the metamaterial's scattering response is recovered for all intermediate angles of incidence.

To clearly underline that the relaxed micromorphic model is able to unveil the complete anisotropic behaviour of the considered metamaterial, we show in this section the scattering solution obtained for two intermediate angles, namely $\theta=15$\textdegree and $\theta=30$\textdegree.
However, for the interested reader, we also report in the Appendix~\ref{appendix-sec1} some particularly interesting simulation solutions obtained for $\theta=0$\textdegree.

Since the incident wave comes from the external Cauchy medium, we are able to set the incident wave to be either a pressure or a shear incident wave. In formulas we have
\begin{align}
    u^{I}(x_1,x_2,t)
    =
    a^{P/S,I} \,
    \psi^{P/S,I} \, e^{i \left( k_1^{P/S,I} \, x_1 + k_2^{P/S,I} \, x_2 - \omega \, t \right)} \, ,
    \,\,
    \text{with}
    \,\,
    \left\{
    \begin{array}{ll}
    k_1^{L,I}
    =
    \sqrt{\dfrac{\rho_{ \text{Al}}}{\lambda_{ \text{Al}} + 2\mu_{ \text{Al}}}\omega^2-\left( k_2^{P,I} \right)^2}
    \\*[5mm]
    k_1^{S,I}
    =
    \sqrt{\dfrac{\rho_{ \text{Al}}}{\mu_{ \text{Al}}}\omega^2-\left( k_2^{S,I} \right)^2}
    \end{array}
    \right.
    \, ,
    \label{eq:plane_wave_plus_k}
\end{align}
where $P/S$ is related to the pressure or shear wave, $I$ means incident, $a^{P/S}$ is the amplitude, $\psi^{P/S,I}$ is the normalized eigenvector associated with the pressure or shear wave, $k_1^{P/S,I}$ and $k_2^{P/S,I}$ are the components of the wave vector, and $\omega$ is the frequency.
The amplitude, frequency and second component of the wave vector are all assigned, while the first component of the wave vector is computed accordingly to eq.(\ref{eq:plane_wave_plus_k})$_2$ which gives the classical dispersion relations for Cauchy continua.

Finally, in order to show the performances of the relaxed micromorphic model for the whole considered frequency range (up to the upper band-gap limit), we also implement a reference finite-element simulation in which the internal region is occupied by a classical Cauchy continuum whose elastic parameters are obtained as the \textit{long-wave} limit of the relaxed micromorphic model and are given in Table~\ref{tab:parameters_RM}.
This configuration is schematically represented in Fig.\ref{fig:geometry}(\textit{left}) and the associated finite-element simulation will show that the correct framework to deal with the medium-high frequency homogenized metamaterial's response is given by the relaxed micromorphic model.
Clearly, in this case, the implemented boundary value problem is given by eq.(\ref{eq:equiCau}) and eq.(\ref{eq:sigClass}).

\begin{figure}[H]
	\centering
	\begin{minipage}[H]{0.32\textwidth}
		\includegraphics[width=\textwidth]{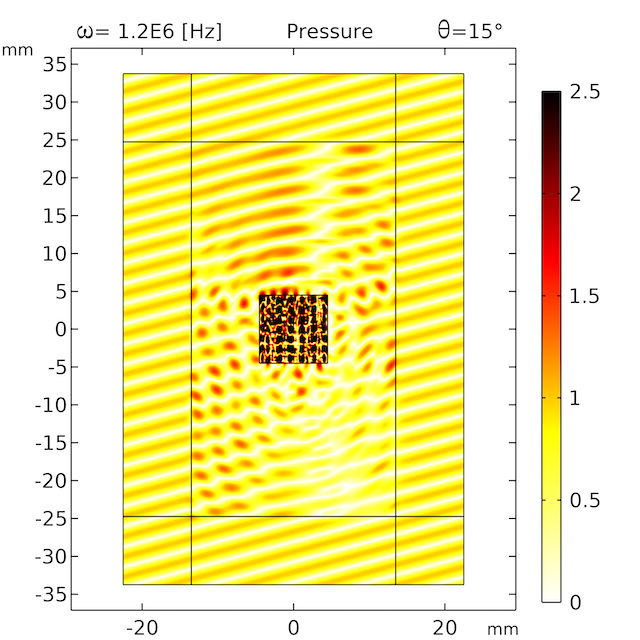}
	\end{minipage}
	\begin{minipage}[H]{0.32\textwidth}
		\includegraphics[width=\textwidth]{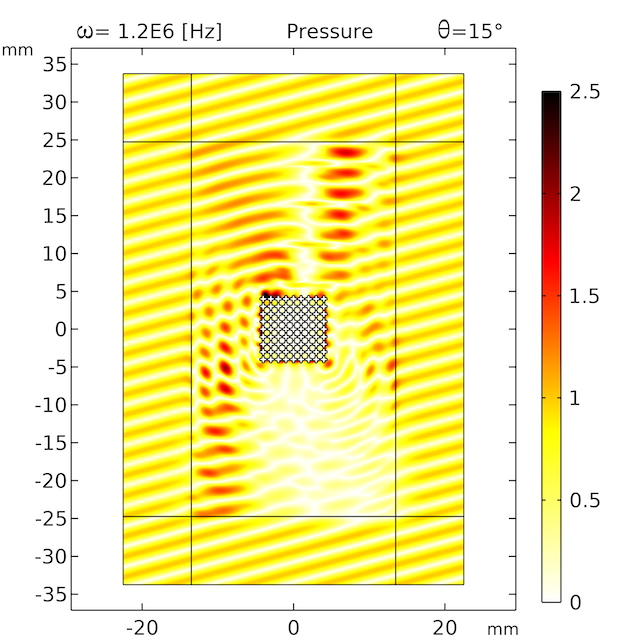}
	\end{minipage}
	\begin{minipage}[H]{0.32\textwidth}
		\includegraphics[width=\textwidth]{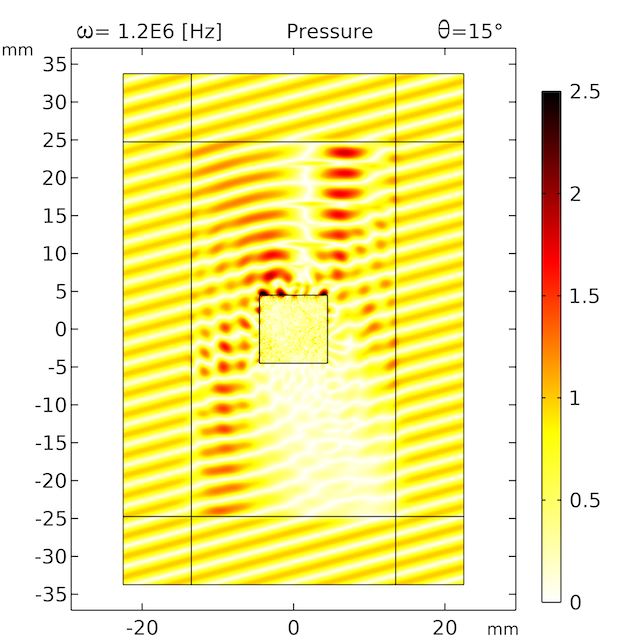}
	\end{minipage}
	\\
	\begin{minipage}[H]{0.32\textwidth}
		\includegraphics[width=\textwidth]{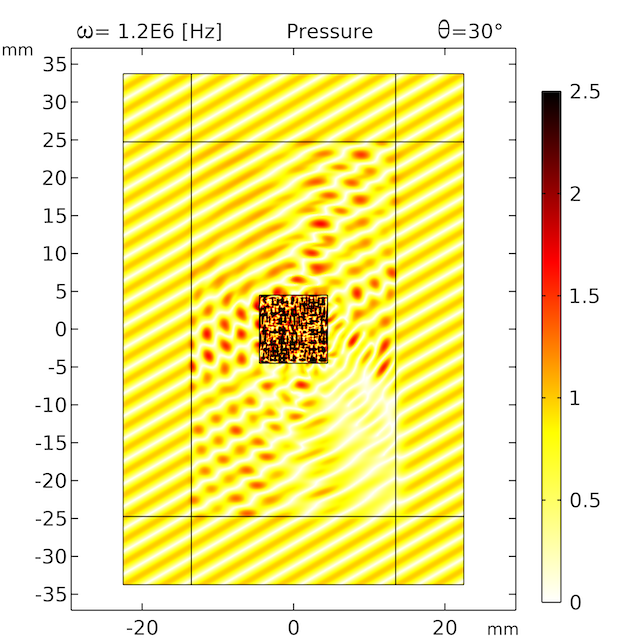}
	\end{minipage}
	\begin{minipage}[H]{0.32\textwidth}
		\includegraphics[width=\textwidth]{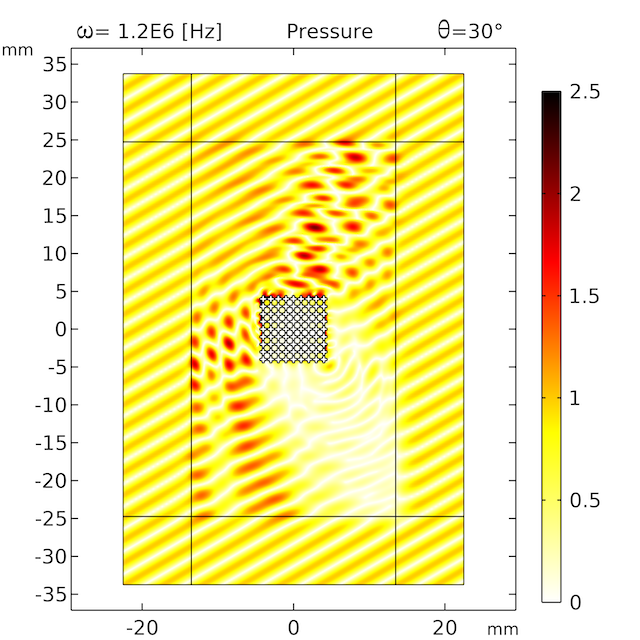}
	\end{minipage}
	\begin{minipage}[H]{0.32\textwidth}
		\includegraphics[width=\textwidth]{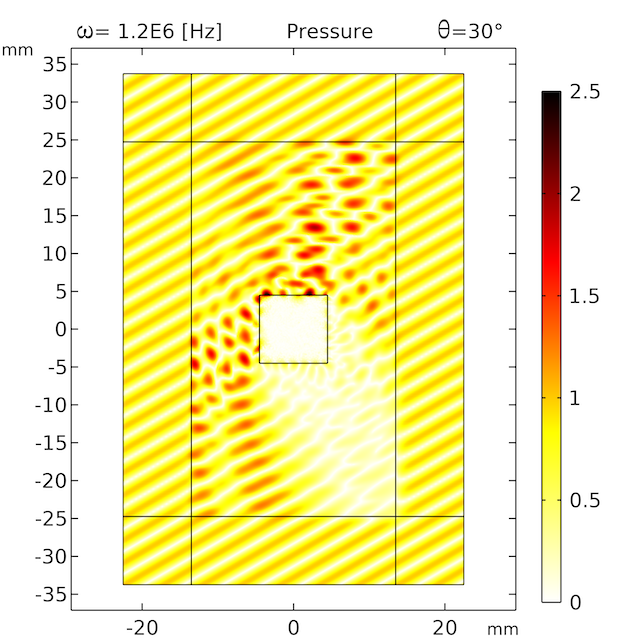}
	\end{minipage}
	\caption{Norm of the dimensionless displacement field (scaled with respect to the amplitude of the incident wave) for a time harmonic incident pressure wave whose frequency is 1.2E6~[Hz] and for two incidence angles of 15 and 30 degrees with respect to the vertical axis.
	(\textit{left}) The metamaterial is modelled with the classical Cauchy model; 
	(\textit{center}) the metamaterial is modelled encoding all the geometrical details; 
	(\textit{right}) the metamaterial is modelled with the relaxed micromorphic material. 
	}
	\label{fig:Cross_compa_p_1}
\end{figure}

\begin{figure}[H]
	\centering
	\begin{minipage}[H]{0.32\textwidth}
		\includegraphics[width=\textwidth]{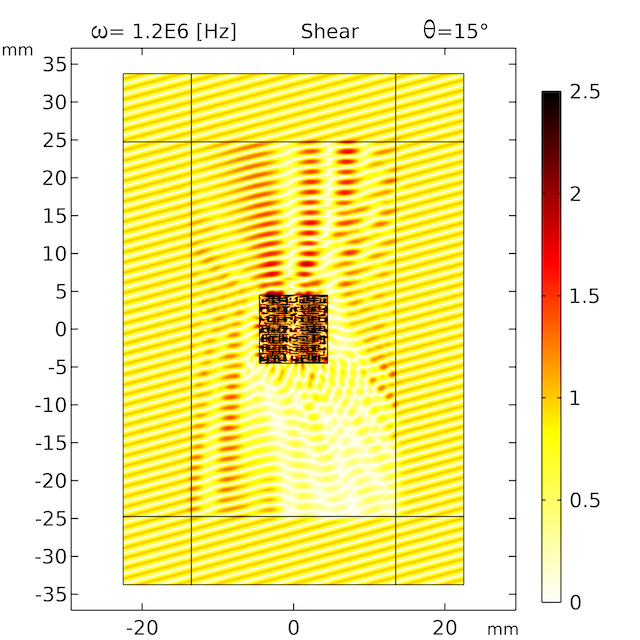}
	\end{minipage}
	\begin{minipage}[H]{0.32\textwidth}
		\includegraphics[width=\textwidth]{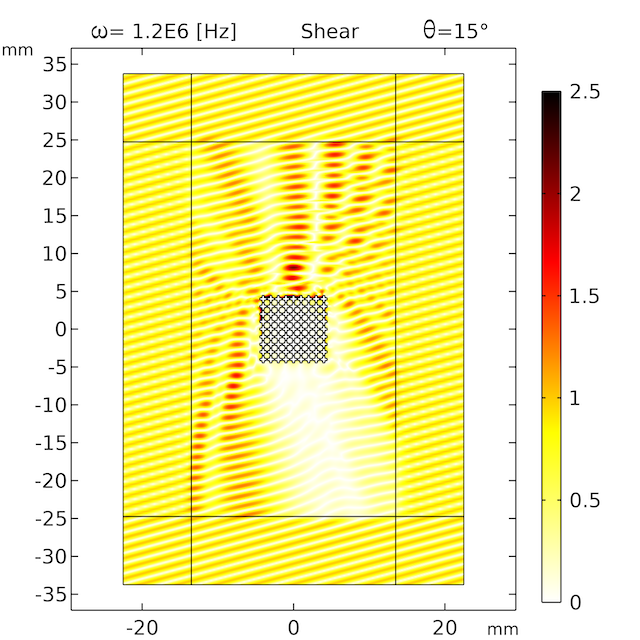}
	\end{minipage}
	\begin{minipage}[H]{0.32\textwidth}
		\includegraphics[width=\textwidth]{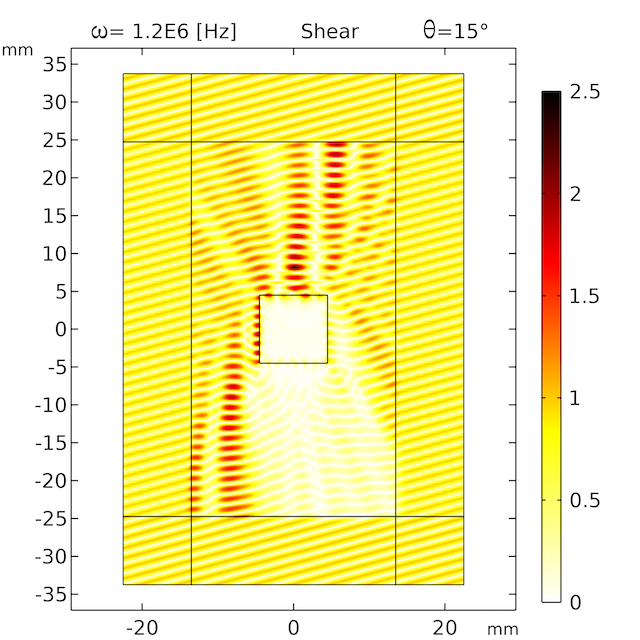}
	\end{minipage}
	\\
	\begin{minipage}[H]{0.32\textwidth}
		\includegraphics[width=\textwidth]{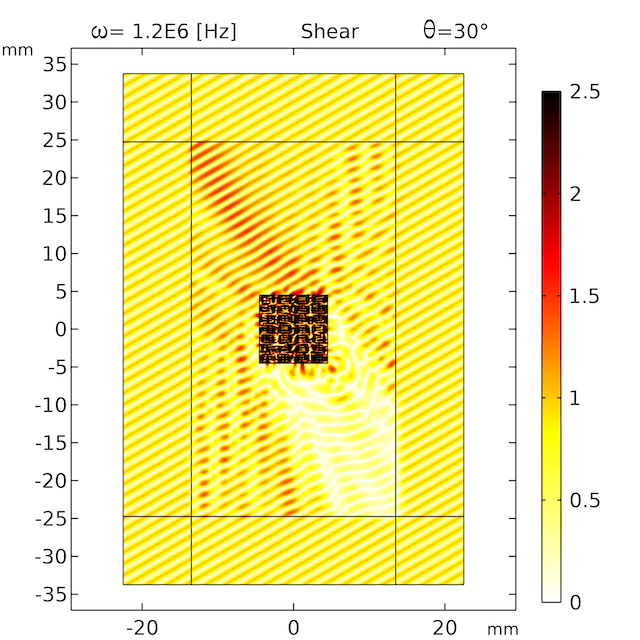}
	\end{minipage}
	\begin{minipage}[H]{0.32\textwidth}
		\includegraphics[width=\textwidth]{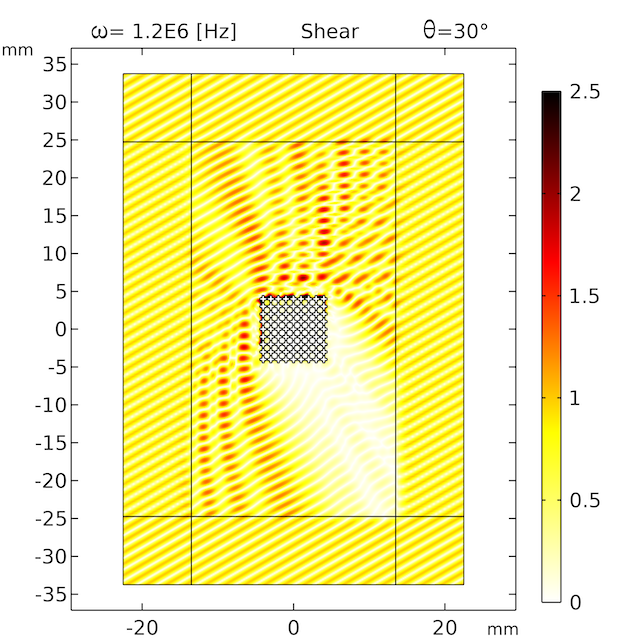}
	\end{minipage}
	\begin{minipage}[H]{0.32\textwidth}
		\includegraphics[width=\textwidth]{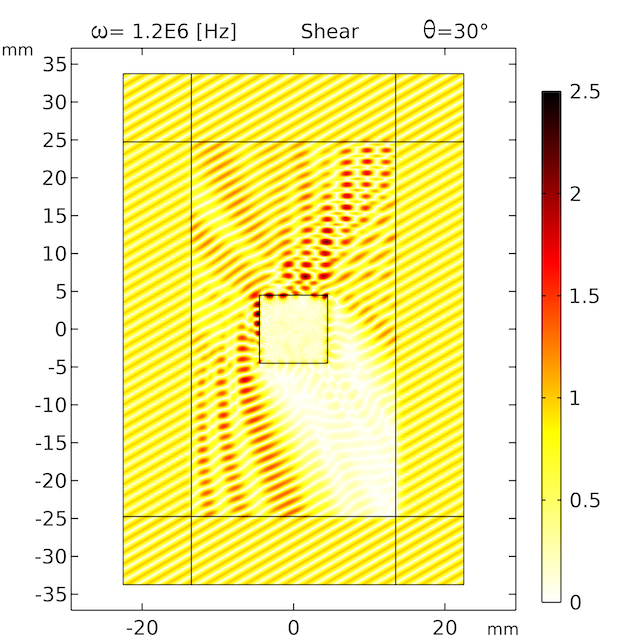}
	\end{minipage}
	\caption{Norm of the dimensionless displacement field (scaled with respect to the amplitude of the incident wave) for a time harmonic incident shear wave whose frequency is 1.2E6~[Hz] and for two incidence angles of 15 and 30 degrees with respect to the vertical axis.
	(\textit{left}) The metamaterial is modelled with the classical Cauchy model; 
	(\textit{center}) the metamaterial is modelled encoding all the geometrical details; 
	(\textit{right}) the metamaterial is modelled with the relaxed micromorphic material. 
	}
	\label{fig:Cross_compa_s_1}
\end{figure}

\begin{figure}[H]
	\centering
	\begin{minipage}[H]{0.32\textwidth}
		\includegraphics[width=\textwidth]{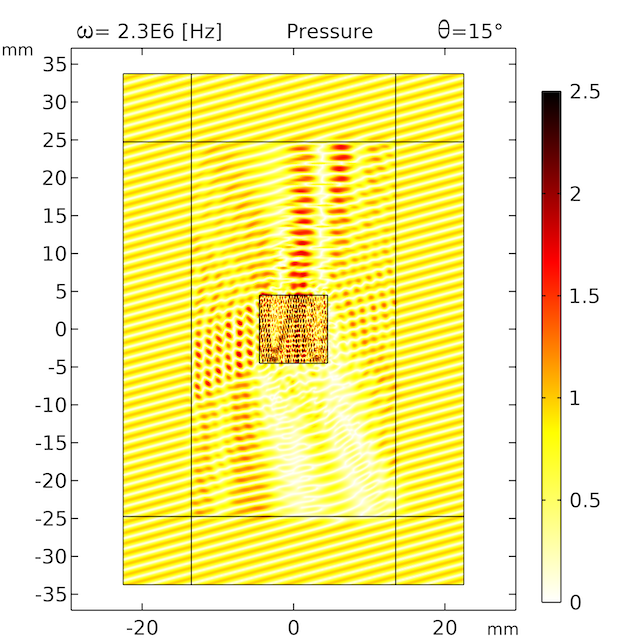}
	\end{minipage}
	\begin{minipage}[H]{0.32\textwidth}
		\includegraphics[width=\textwidth]{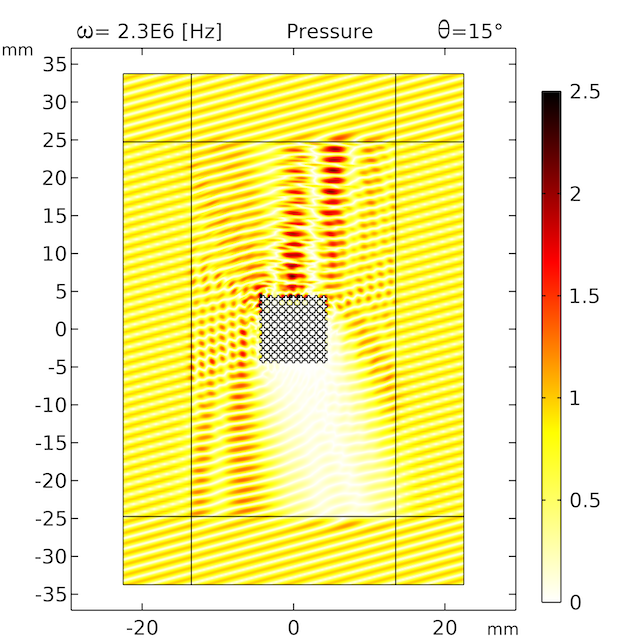}
	\end{minipage}
	\begin{minipage}[H]{0.32\textwidth}
		\includegraphics[width=\textwidth]{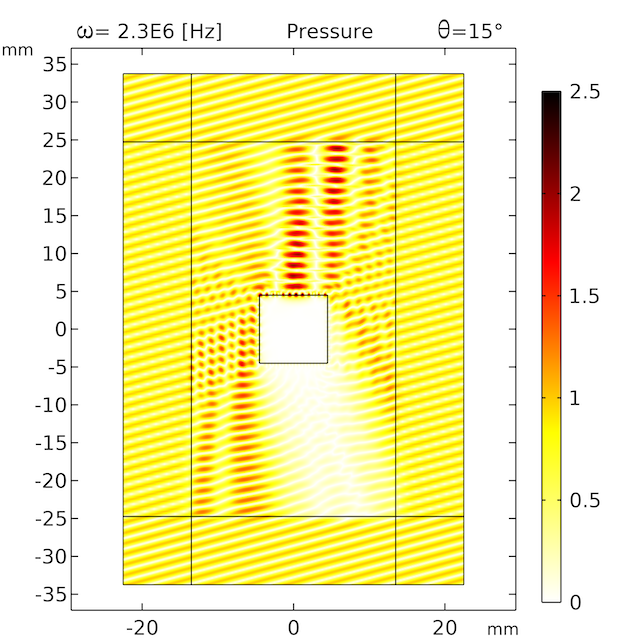}
	\end{minipage}
	\\
	\begin{minipage}[H]{0.32\textwidth}
		\includegraphics[width=\textwidth]{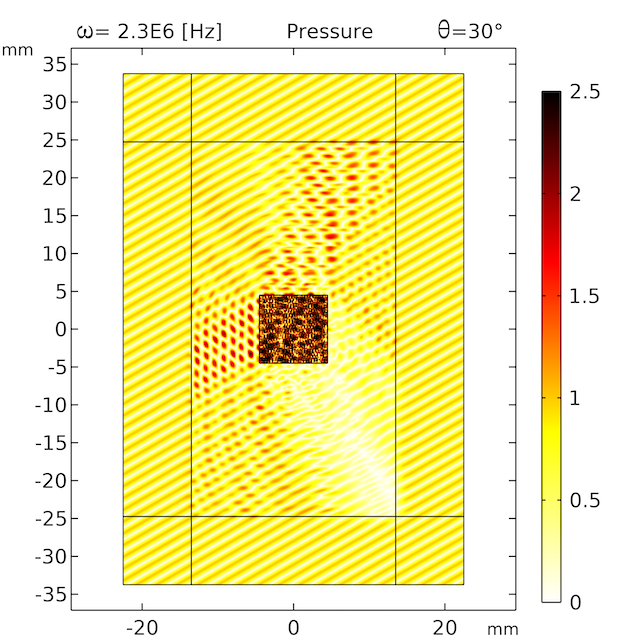}
	\end{minipage}
	\begin{minipage}[H]{0.32\textwidth}
		\includegraphics[width=\textwidth]{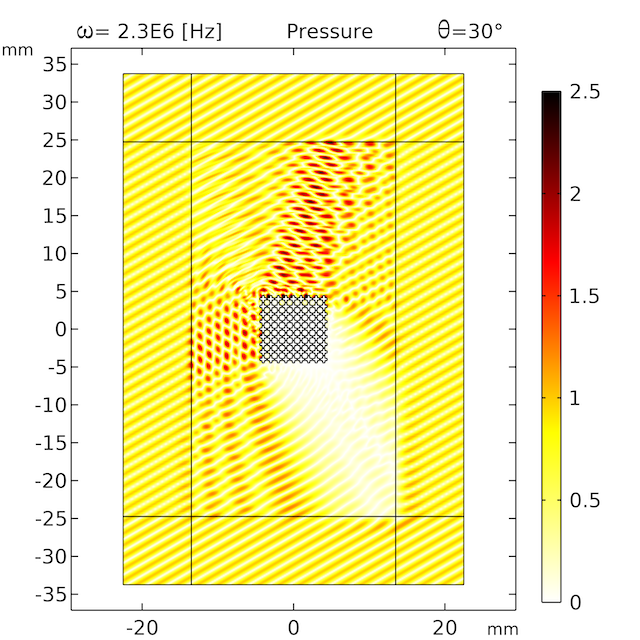}
	\end{minipage}
	\begin{minipage}[H]{0.32\textwidth}
		\includegraphics[width=\textwidth]{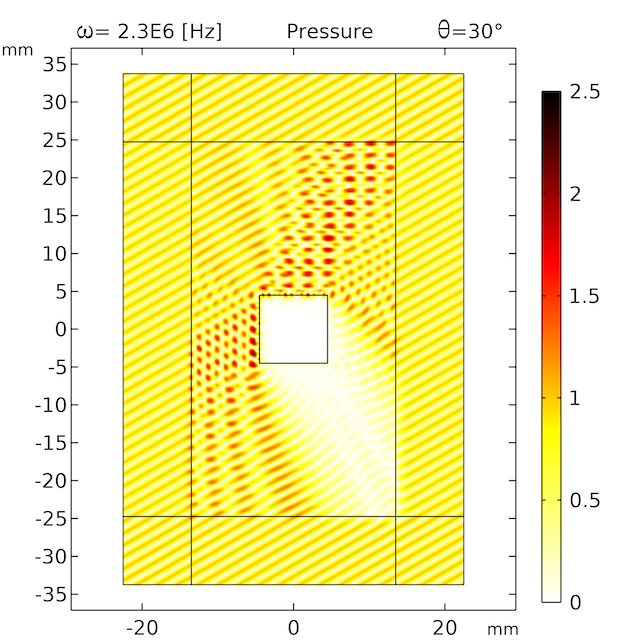}
	\end{minipage}
	\caption{Norm of the dimensionless displacement field (scaled with respect to the amplitude of the incident wave) for a time harmonic incident pressure wave whose frequency is 2.3E6~[Hz] and for two incidence angles of 15 and 30 degrees with respect to the vertical axis.
	(\textit{left}) The metamaterial is modelled with the classical Cauchy model; 
	(\textit{center}) the metamaterial is modelled encoding all the geometrical details; 
	(\textit{right}) the metamaterial is modelled with the relaxed micromorphic material. 
	}
	\label{fig:Cross_compa_p_2}
\end{figure}

\begin{figure}[H]
	\centering
	\begin{minipage}[H]{0.32\textwidth}
		\includegraphics[width=\textwidth]{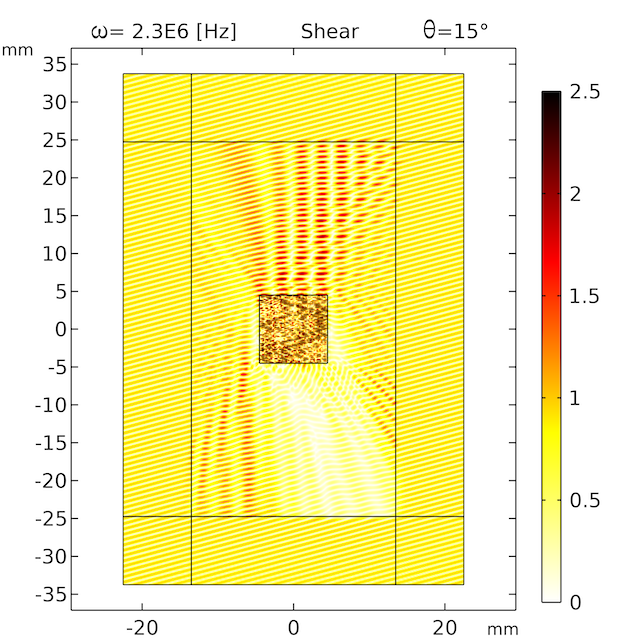}
	\end{minipage}
	\begin{minipage}[H]{0.32\textwidth}
		\includegraphics[width=\textwidth]{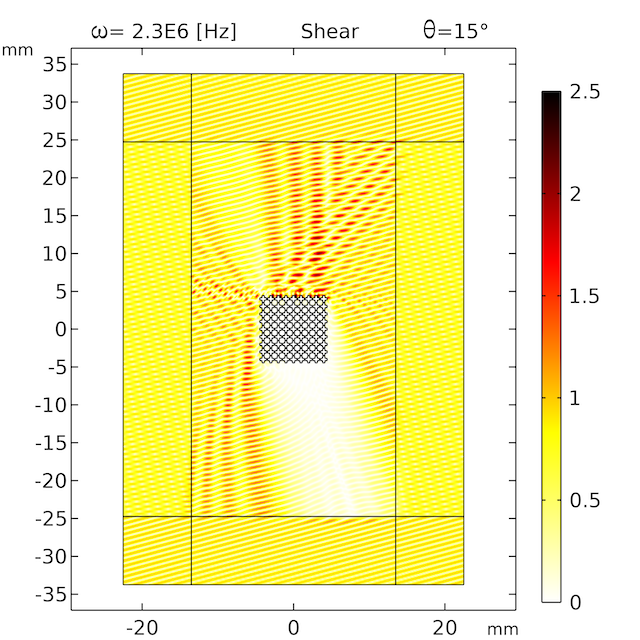}
	\end{minipage}
	\begin{minipage}[H]{0.32\textwidth}
		\includegraphics[width=\textwidth]{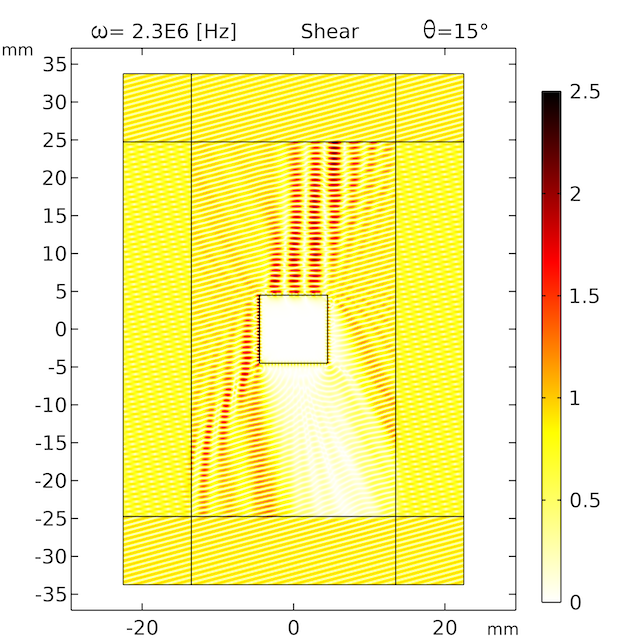}
	\end{minipage}
	\\
	\begin{minipage}[H]{0.32\textwidth}
		\includegraphics[width=\textwidth]{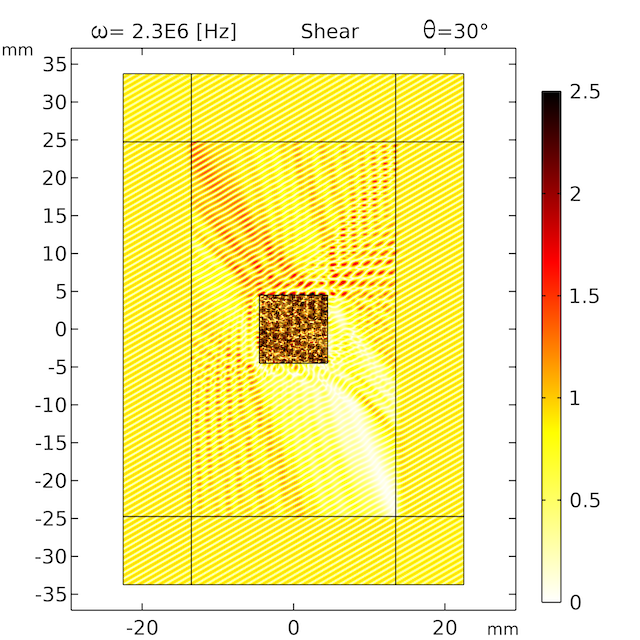}
	\end{minipage}
	\begin{minipage}[H]{0.32\textwidth}
		\includegraphics[width=\textwidth]{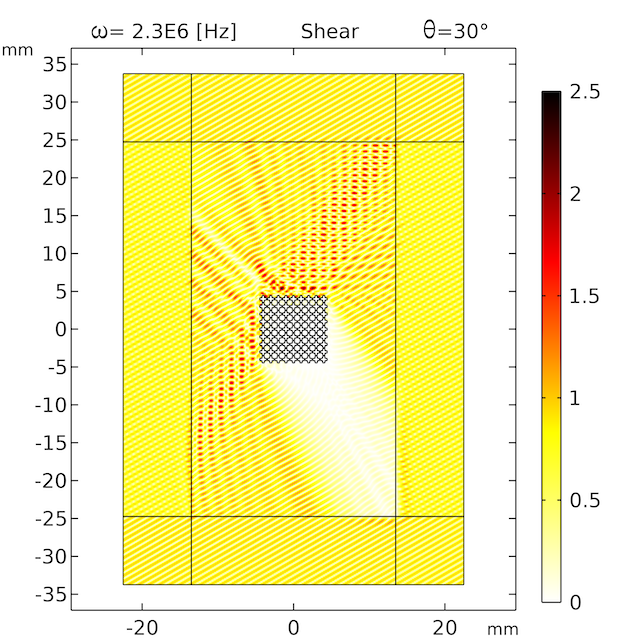}
	\end{minipage}
	\begin{minipage}[H]{0.32\textwidth}
		\includegraphics[width=\textwidth]{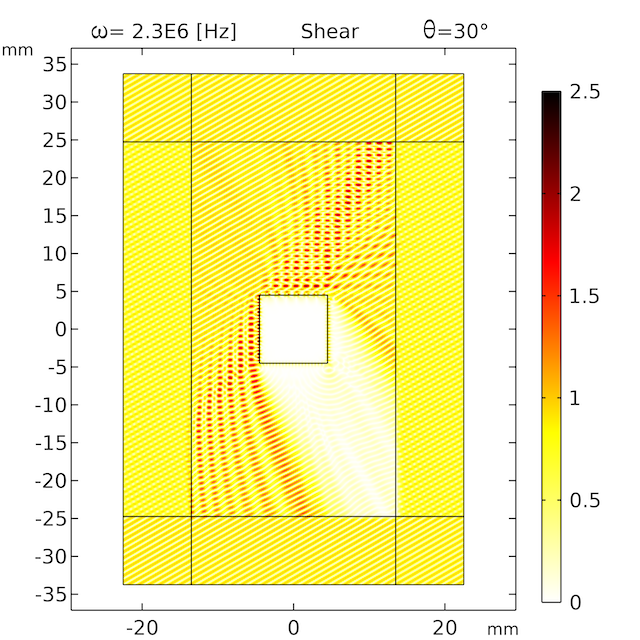}
	\end{minipage}
	\caption{Norm of the dimensionless displacement field (scaled with respect to the amplitude of the incident wave) for a time harmonic incident shear wave whose frequency is 2.3E6~[Hz] and for two incidence angles of 15 and 30 degrees with respect to the vertical axis.
	(\textit{left}) The metamaterial is modelled with the classical Cauchy model; 
	(\textit{center}) the metamaterial is modelled encoding all the geometrical details; 
	(\textit{right}) the metamaterial is modelled with the relaxed micromorphic material. 
	}
	\label{fig:Cross_compa_s_2}
\end{figure}

Figures~\ref{fig:Cross_compa_p_1}-\ref{fig:Cross_compa_s_2} show the metamaterial's scattering response, as obtained with the equivalent Cauchy (\textit{left}), the micro-structured (\textit{center}) and the relaxed micromorphic simulations (\textit{right}), for the two considered angles of incidence and for two frequency laying in the metamaterial's band-gap region (lower band-gap and higher band-gap regions).
By direct inspection of Fig.~\ref{fig:Cross_compa_p_1} and Fig.~\ref{fig:Cross_compa_s_1}, it is possible to notice that the relaxed micromorphic model correctly describes the metamaterial's scattering response for all considered frequencies, angles of incidence, and for both pressure and shear incidence waves.
On the other hand, it can be directly inferred from these figures that a classical Cauchy continuum model is not able to describe the metamaterial's scattering patterns at the considered band-gap frequency.

Similar results can be retrieved in Fig.~\ref{fig:Cross_compa_p_2} and Fig.~\ref{fig:Cross_compa_s_2} that are relative to a higher frequency value taken in the upper band-gap region.
\section{Shield device metastructure allowing energy focusing}
\begin{figure}[H]
	\centering
	\begin{minipage}[H]{0.45\textwidth}
	\centering
	\includegraphics[width=0.9\textwidth]{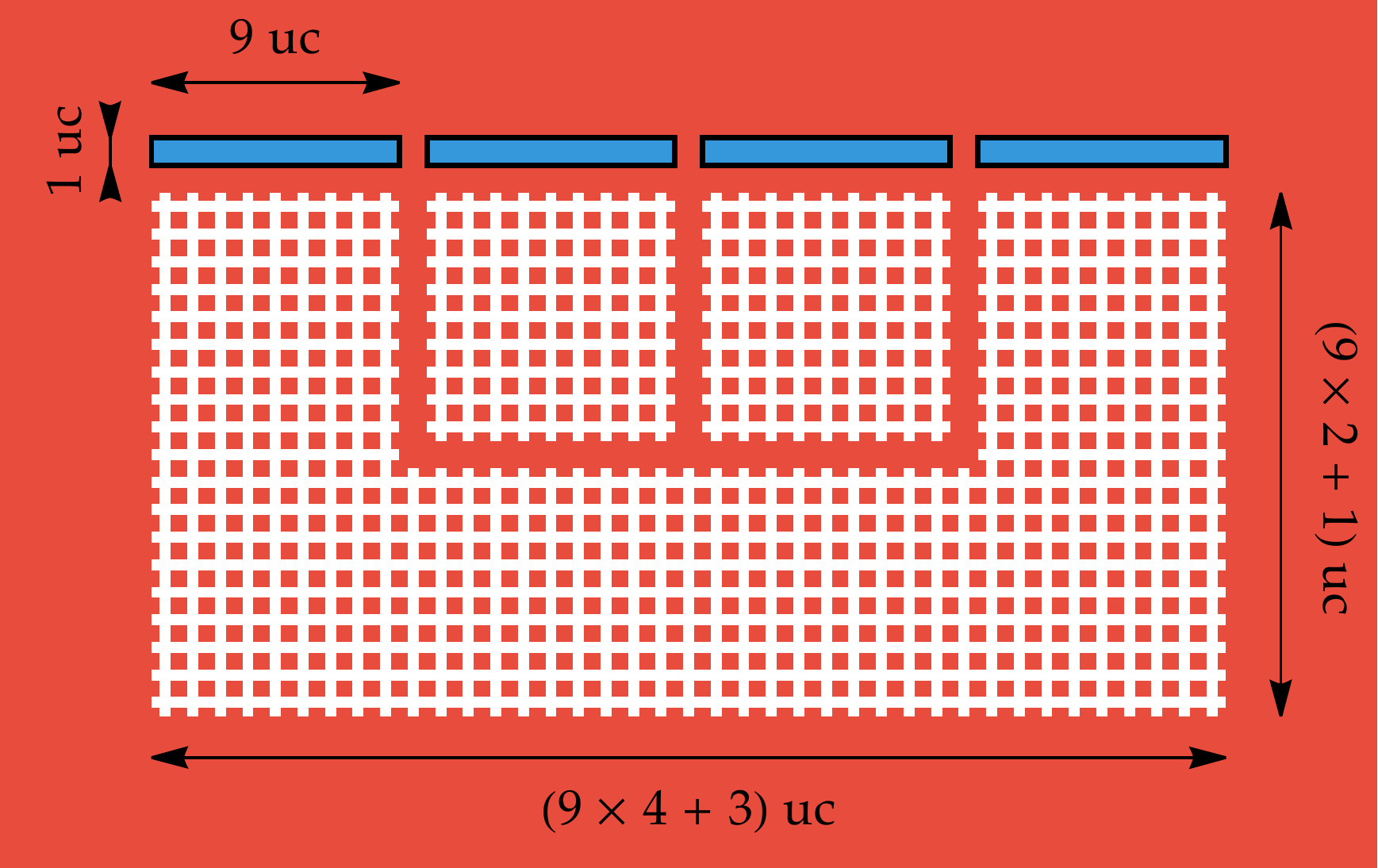}
	\end{minipage}
	\hfill
	\begin{minipage}[H]{0.45\textwidth}
	\includegraphics[width=0.9\textwidth]{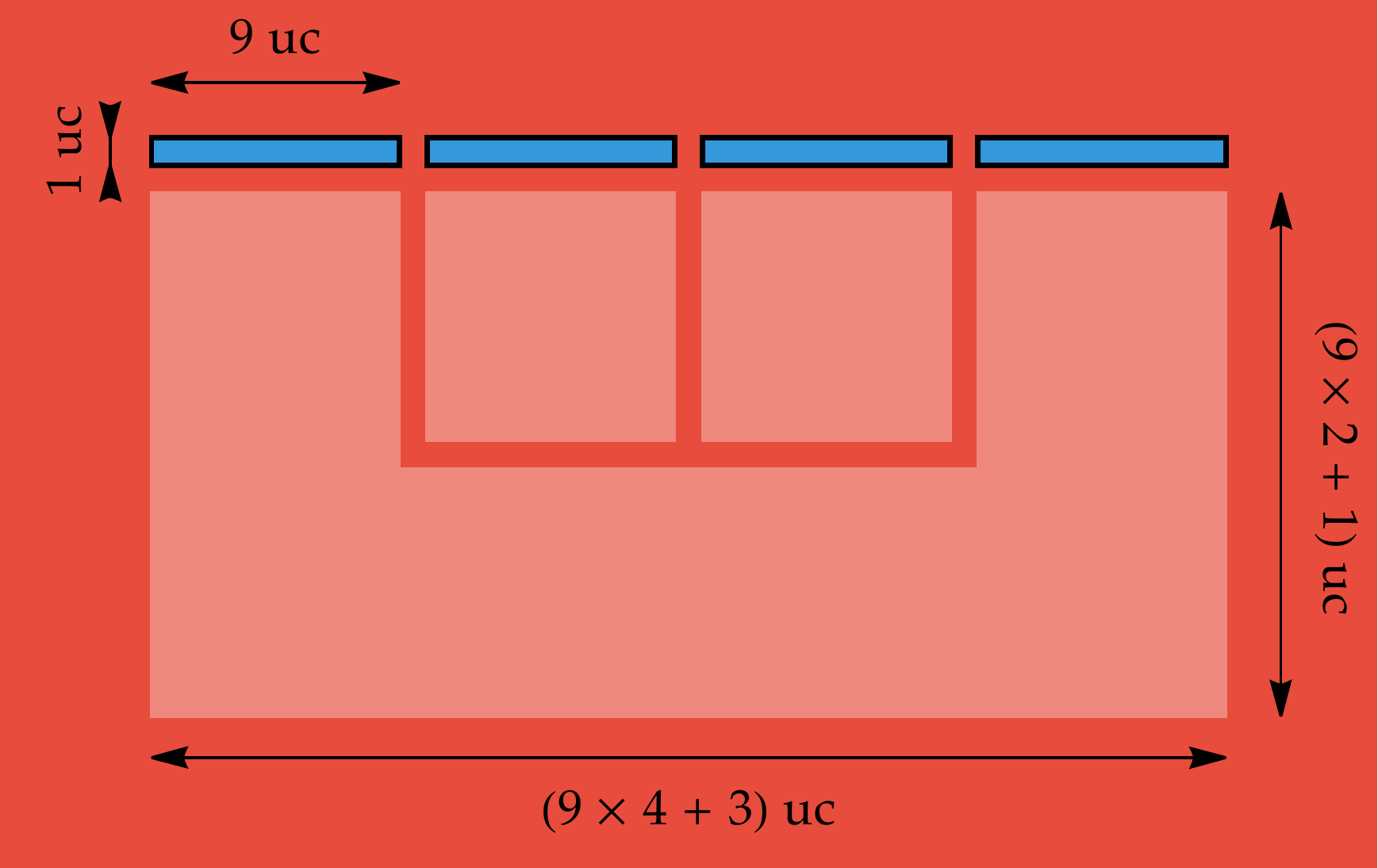}
	\end{minipage}
	\caption{Scheme of (\textit{left}) the metastructure and (\textit{right}) the equivalent relaxed micromorphic structure. The continuous red region is made-up of a Cauchy continuum whose elastic properties are given in Table~\ref{tab:unit_cell}.
	The metamaterial's unit cell in the left panel is given in Fig.~\ref{fig:unit_cell} and the relaxed micromorphic elastic coefficients used to model the metamaterial in the right panel are given in Table~\ref{tab:parameters_RM}.}
	\label{fig:meta_structure_scheme}
\end{figure}

Once the relaxed micromorphic boundary value problem has been validated to describe the scattering of a finite-size metamaterial, we exploit the computational performances of the relaxed micromorphic model to explore a complex metastructure that can act as a protection device, while focusing energy for its eventual subsequent reuse. 
We present here the metastructure in Fig.~\ref{fig:meta_structure_scheme}, whose intent is to channel and focus the energy of the incident wave along specific paths with the aim of harvesting such energy, while, in addition, also acts as a shield.

The continuity conditions enforced at the interfaces between the external Cauchy medium and the equivalent relaxed micromorphic medium are the same reported in eqs.(\ref{eq:jump_cond_2}), where $\partial \Omega$ is composed of all the boundary delimiting the domains made-up of relaxed micromorphic material.

\begin{figure}[H]
	\centering
	\begin{minipage}[H]{0.45\textwidth}
		\includegraphics[width=\textwidth]{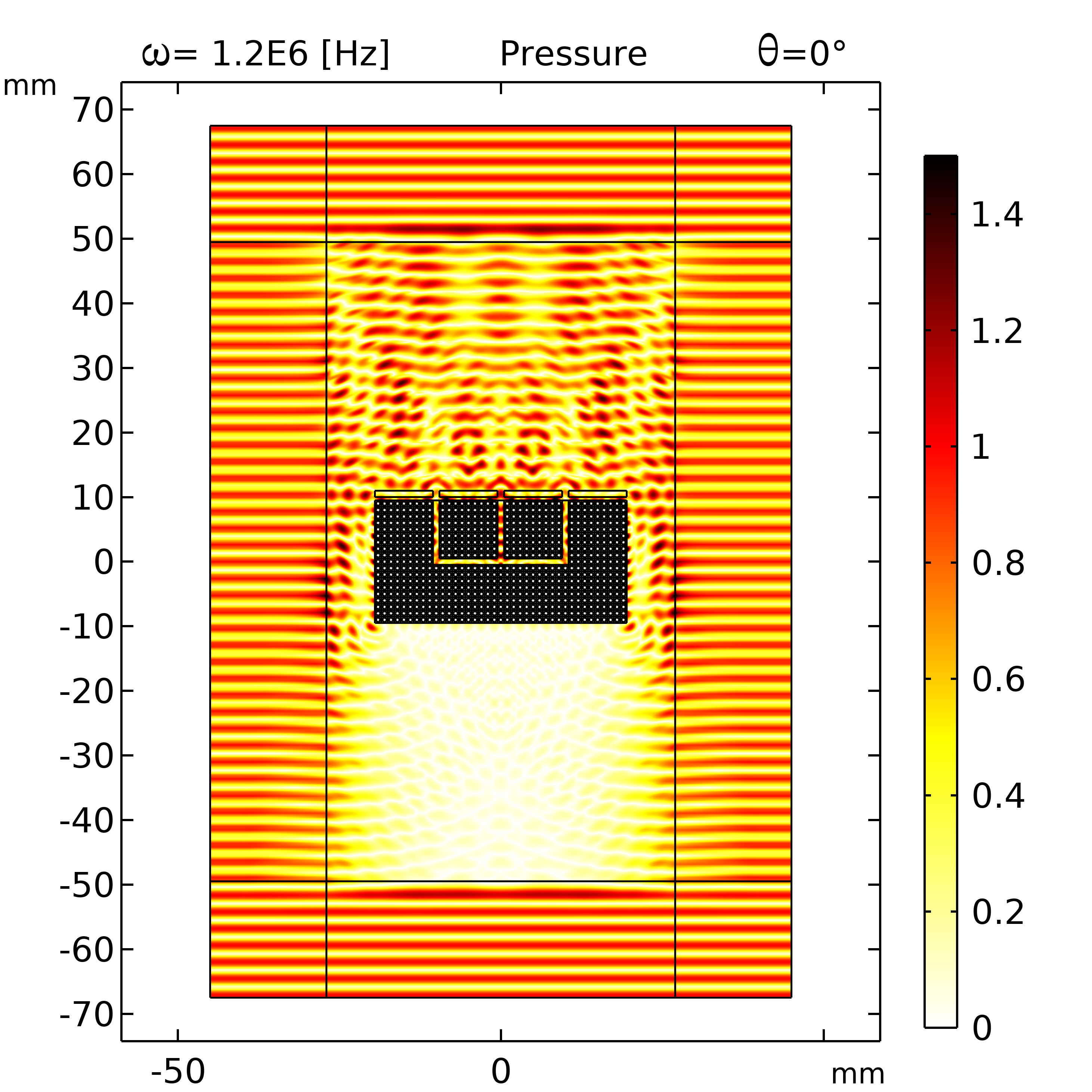}
	\end{minipage}
	\hfill
	\begin{minipage}[H]{0.45\textwidth}
		\includegraphics[width=\textwidth]{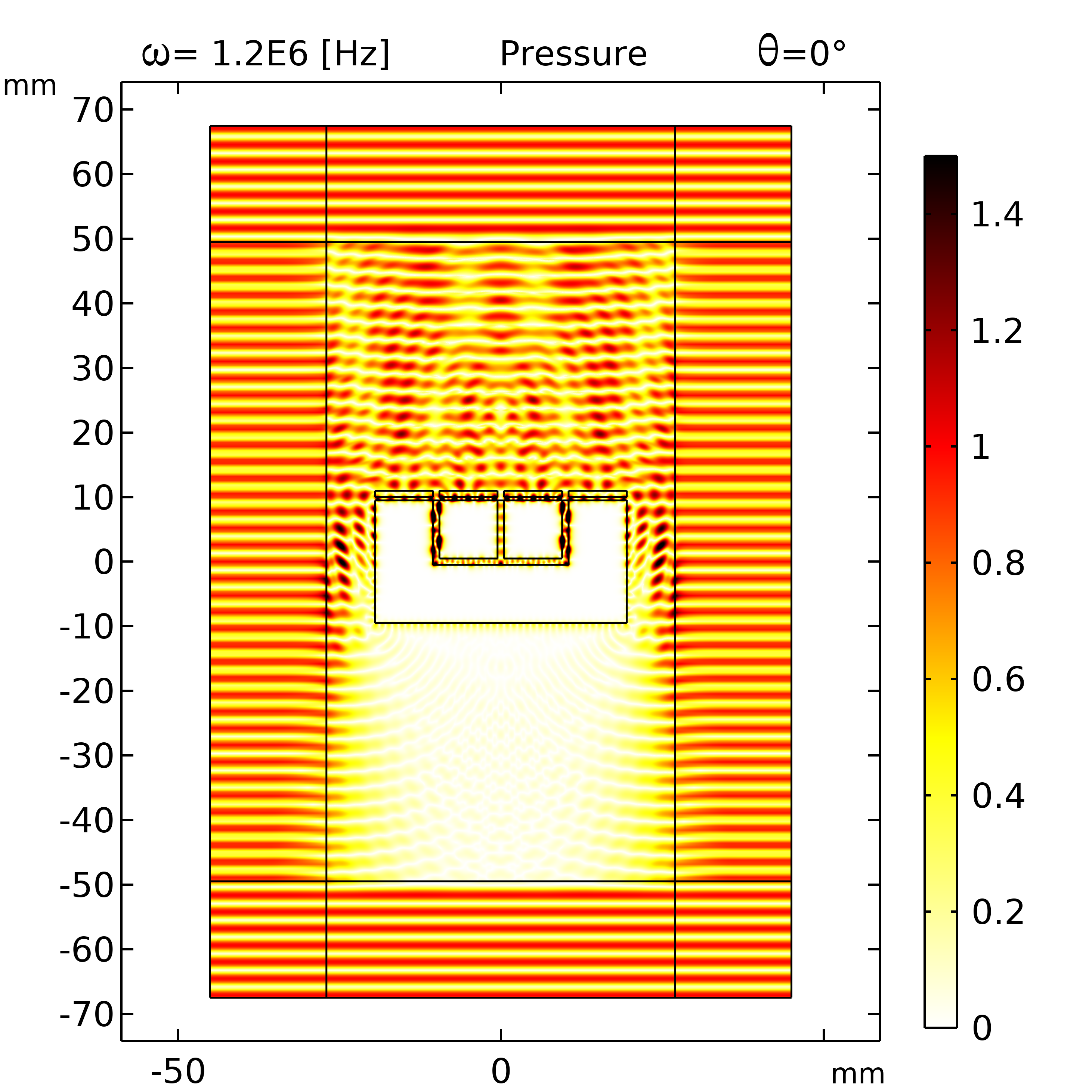}
	\end{minipage}
	\caption{Norm of the dimensionless displacement field (scaled with respect to the amplitude of the incident wave) for (\textit{left}) micro-structured material and (\textit{right}) the equivalent relaxed micromorphic material.}
	\label{fig:meta_structure_comparison}
\end{figure}

\begin{figure}[H]
	\centering
	\begin{minipage}[H]{0.45\textwidth}
	\centering
	\includegraphics[width=0.9\textwidth]{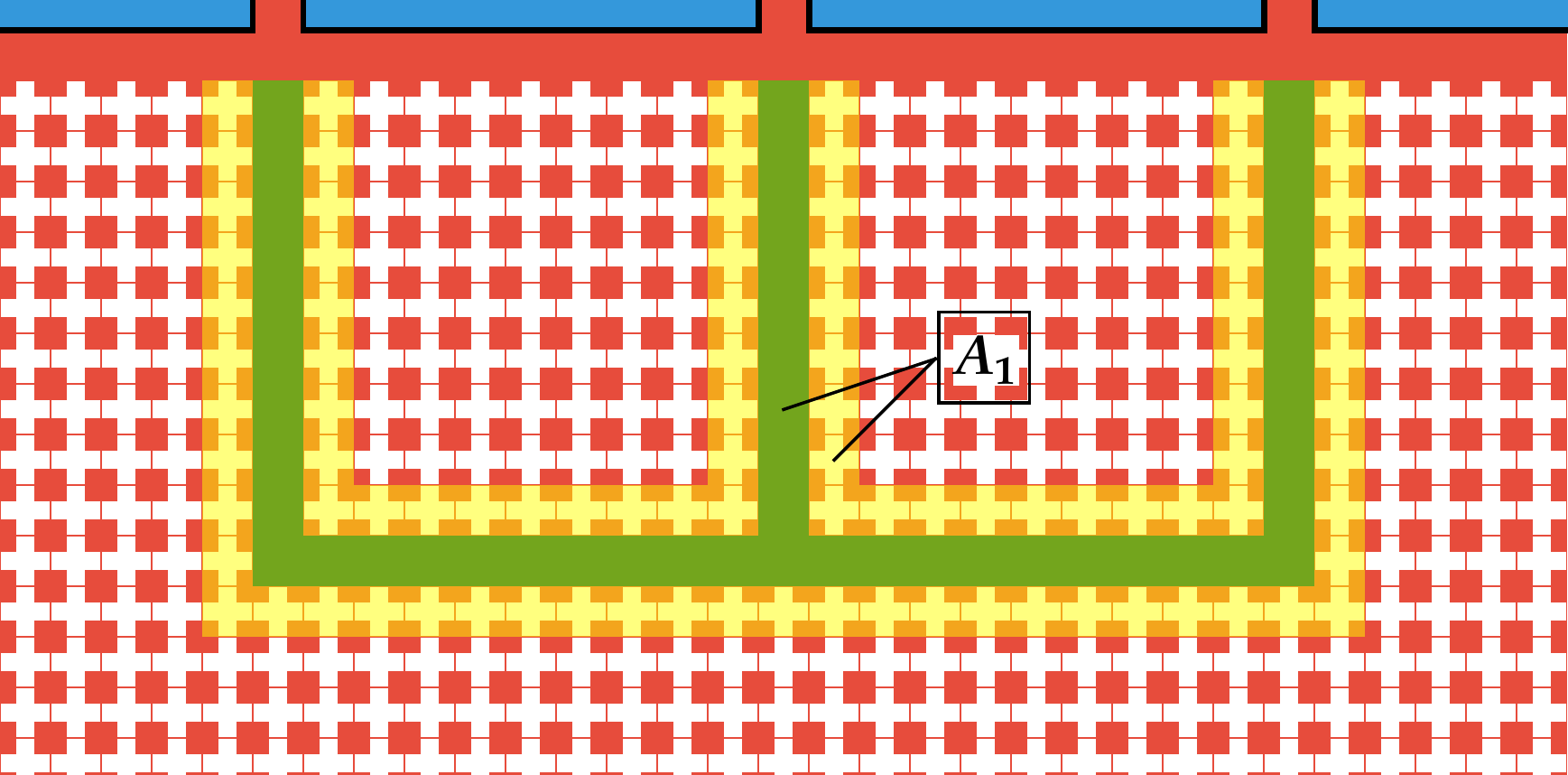}
	\end{minipage}
	\hfill
	\begin{minipage}[H]{0.45\textwidth}
	\includegraphics[width=0.9\textwidth]{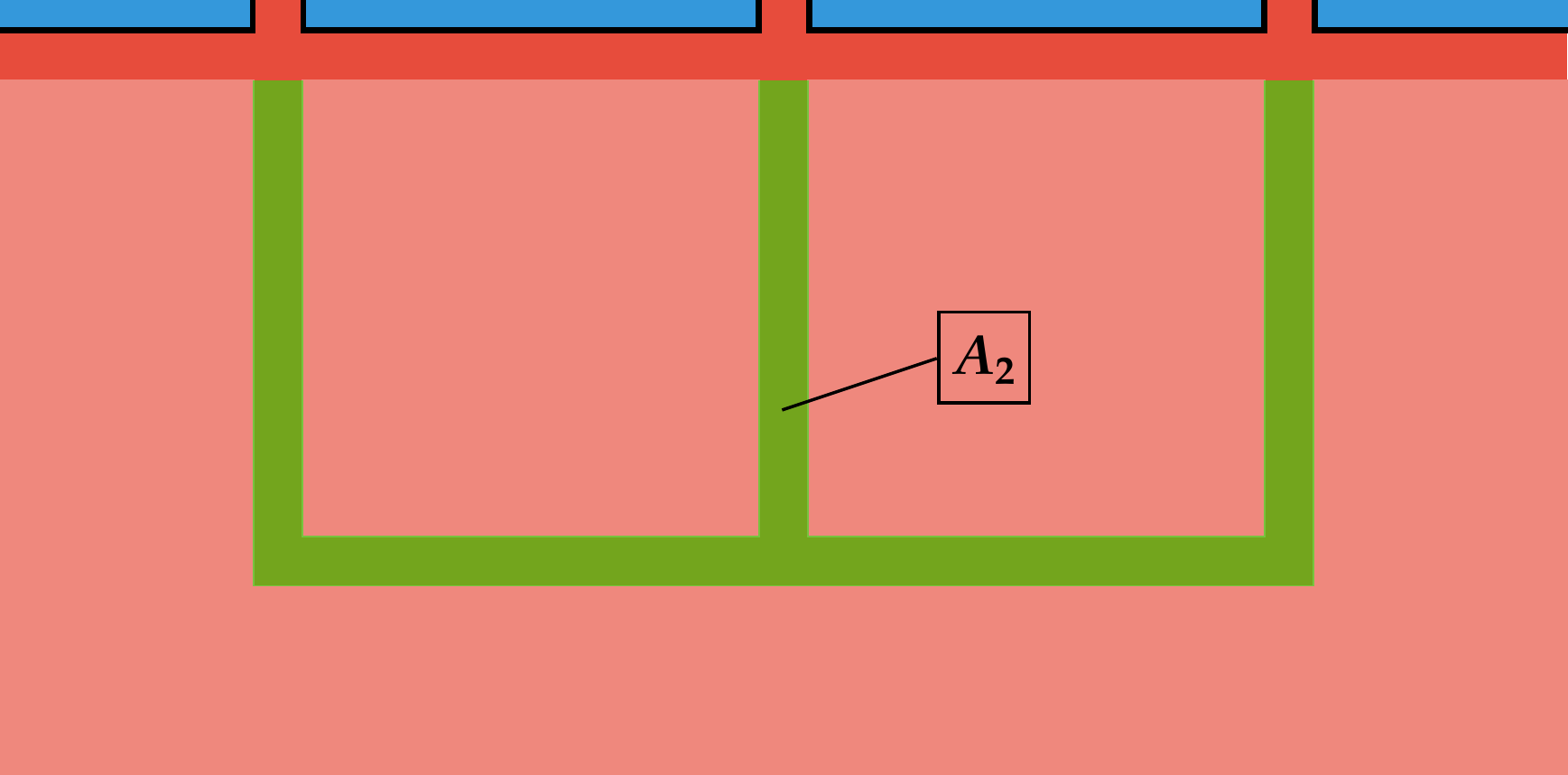}
	\end{minipage}
	\caption{Detail of the domain considered for the evaluation of the average energy density for (\textit{left}) the metastructure and (\textit{right}) the equivalent structure  made-up of the equivalent relaxed micromorphic material.}
	\label{fig:meta_structure_scheme_2}
\end{figure}

This metastructure is subjected to a downward time harmonic pressure wave (1.2E6~[Hz]) which propagates parallel to the vertical axis.
The blue rectangles in Fig.~\ref{fig:meta_structure_scheme} represent dumpers, which are designed to reduce the reflected wave in the higher part of the domain.
In Fig.~\ref{fig:meta_structure_comparison} is reported the comparison between the norm of the dimensionless displacement field (scaled with respect to the amplitude of the incident wave) for the micro-structured material (\textit{left}) and the equivalent relaxed micromorphic material (\textit{right}).
As a measure of how much energy is focused in the four channels, we evaluate the ratio between the average energy density stored in the channels and the average energy density stored in the outer Cauchy material.
For a quantitative comparison, we take the weighted sum of the green and yellow domain for the micro-structured simulation (Fig.~\ref{fig:meta_structure_scheme_2}(\textit{left})), while only the green portion for the equivalent relaxed micromorphic simulation (Fig.~\ref{fig:meta_structure_scheme_2}(\textit{right})).
This has been done since in the micro-structured material simulation we do not have a neat interface between the domain in which we focus the incident wave and the first layer of unit cells, while we have one in the equivalent relaxed micromorphic material one. This makes the scattered wave focus on the boundary of the cross shaped holes in the yellow region for the micro-structured material simulation, while the energy is focused on the boundary of the green interface when considering the relaxed micromorphic simulation. 
After these considerations, we calculate that the ratio between the average energy density stored in the channels and the average energy density stored in the remaining Cauchy material is approximately 5 and the results given by the two simulations are in quantitative agreement.
In formulas, we can write
\begin{align}
    \frac{W_1}{A_1}
    \approx
    \frac{W_2}{A_2}
    \approx
    5 \,
    \frac{W^{\text{ext}}}{A^{\text{ext}}} \, ,
    \label{eq:energy_fem_ratio}
\end{align}
where $A_1$ is the area of the ``yellow+green'' region, $A_2$ is the area of the ``green'' region, $W_1$ is the weighted energy contained in $A_1$, and $W_2$ is the energy contained in $A_2$.
Moreover, $A^{\text{ext}}$ and $W^{\text{ext}}$ are the area of the outer Cauchy material and the energy stored, respectively, for both the models.
This means that the average concentration of energy is 5 times higher in the channels than outside.
Energy converters can be locally used to recover and reuse the energy initially carried out by the incident wave, and an optimization of the proposed network can be explored to further increase the efficiency of the structure.
\section{Conclusions}
In this paper, we have shown that, once well-posed boundary and interface conditions are established, the relaxed micromorphic model can be efficiently used to model the scattering patterns of finite-size metamaterials. While the relaxed micromorphic model was previously validated on metamaterials' slabs that are finite in one direction, but unbounded in the other \cite{rizzi2020towards,rizzi_exploring_2021,aivaliotis_frequency-_2020}, it is the first time that the model is validated on fully finite-size specimens. This validation holds for a wide range of frequencies and for all possible directions of propagation of the incident wave, thus showing that our model is able to fully unveil the metamaterials' anisotropic responses in the dynamic regime.
Moreover, exploiting the computational performances of the model, we are able to design a metastructure (combining metamaterials and homogeneous Cauchy materials) that acts as a protection device, while focusing energy in specific paths for eventual subsequent conversion and reuse.
In future works, we will build on these results to further optimize metastructures of the type presented here in view of their application for acoustic and vibration control. The present paper thus opens new perspectives for the shot-term design of realistic sustainable metastructures that can control elastic waves while recovering energy.

{\scriptsize
	\paragraph{{\scriptsize Acknowledgements.}}
	Angela Madeo acknowledges support from the European Commission through the funding of the ERC Consolidator Grant META-LEGO, N° 101001759.
	Angela Madeo and Gianluca Rizzi acknowledge funding from the French Research Agency ANR, “METASMART” (ANR-17CE08-0006).
	Angela Madeo and Gianluca Rizzi acknowledge support from IDEXLYON in the framework of the “Programme Investissement d'Avenir” ANR-16-IDEX-0005.
	Patrizio Neff acknowledges support in the framework of the DFG-Priority Programme 2256 ``Variational Methods for Predicting Complex Phenomena in Engineering Structures and Materials'', Neff 902/10-1, Project-No. 440935806.
}


\begingroup
\setstretch{0.8}
\setlength\bibitemsep{3pt}
\printbibliography
\endgroup

\appendix
\section{An example for a normal incident wave for different frequencies}
\label{appendix-sec1}
\begin{figure}[H]
	\centering
	\begin{minipage}[H]{0.32\textwidth}
		\includegraphics[width=\textwidth]{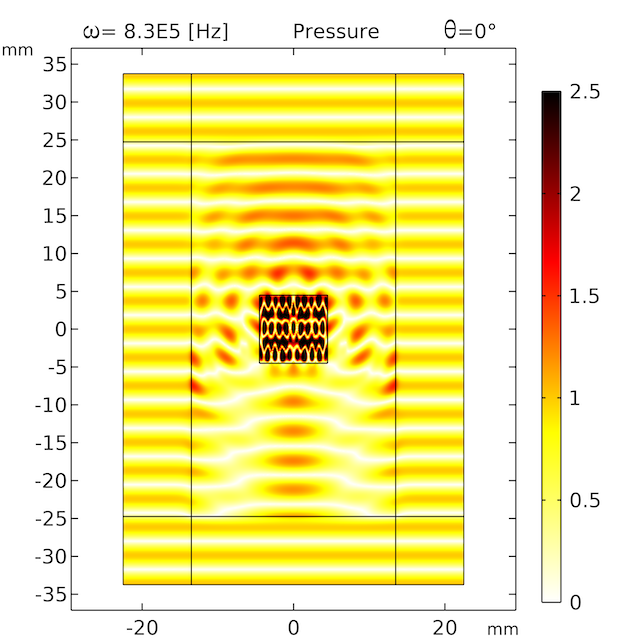}
	\end{minipage}
	\begin{minipage}[H]{0.32\textwidth}
		\includegraphics[width=\textwidth]{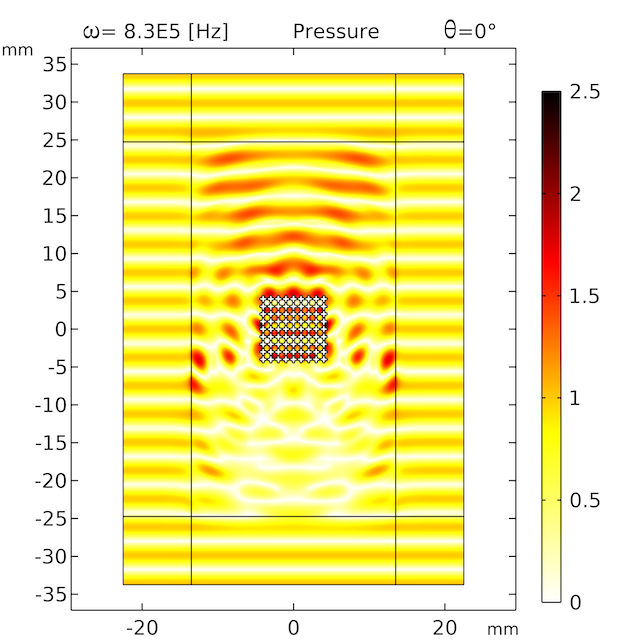}
	\end{minipage}
	\begin{minipage}[H]{0.32\textwidth}
		\includegraphics[width=\textwidth]{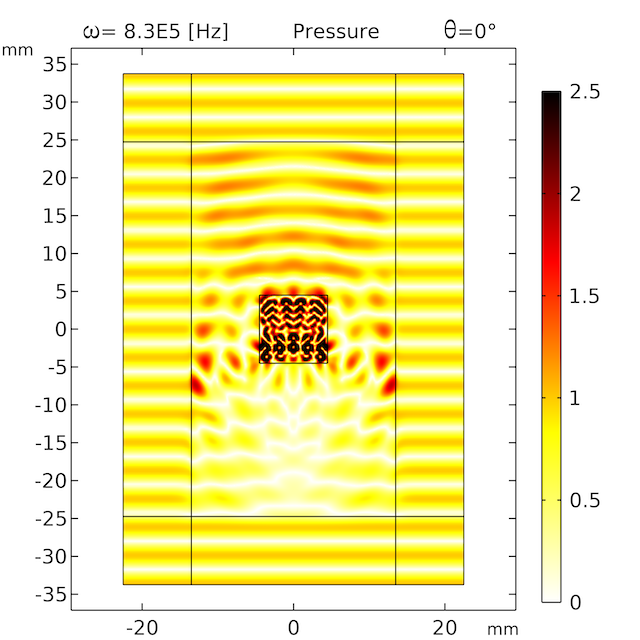}
	\end{minipage}
	\\
	\begin{minipage}[H]{0.32\textwidth}
		\includegraphics[width=\textwidth]{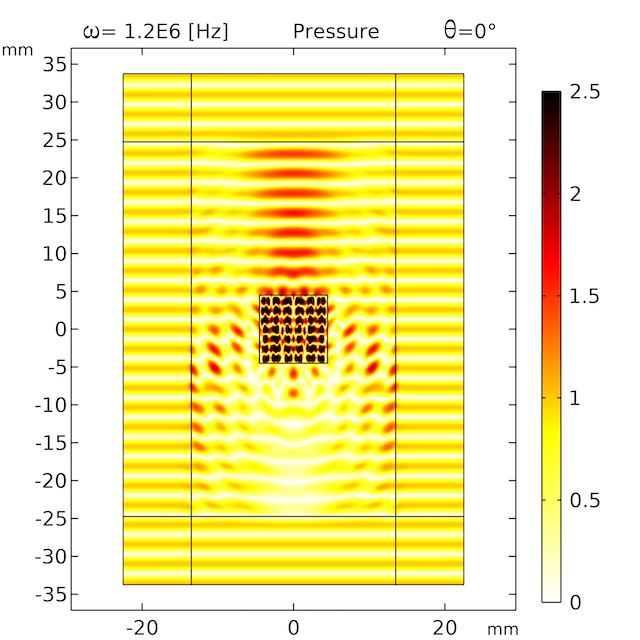}
	\end{minipage}
	\begin{minipage}[H]{0.32\textwidth}
		\includegraphics[width=\textwidth]{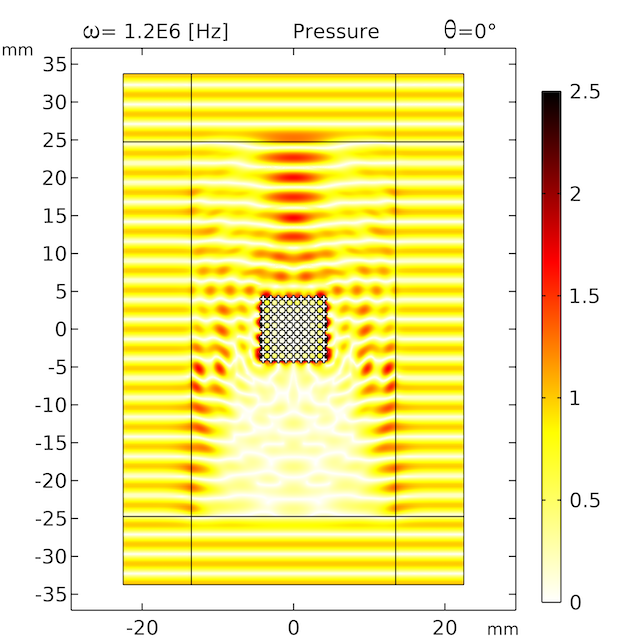}
	\end{minipage}
	\begin{minipage}[H]{0.32\textwidth}
		\includegraphics[width=\textwidth]{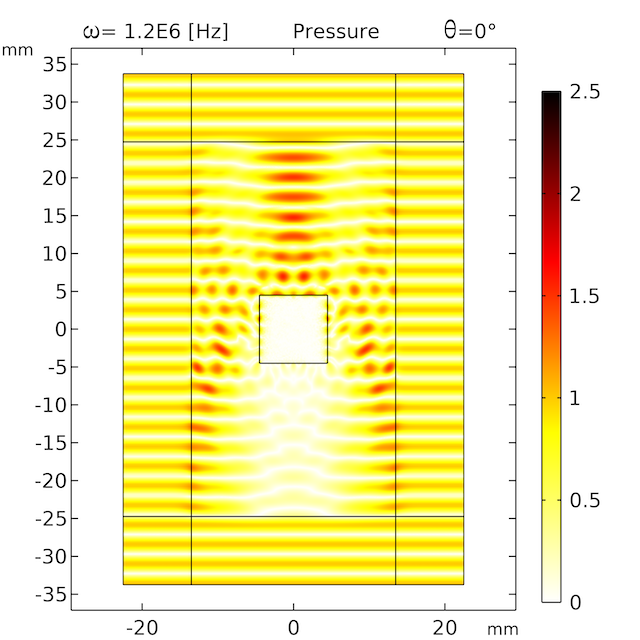}
	\end{minipage}
	\\
	\begin{minipage}[H]{0.32\textwidth}
		\includegraphics[width=\textwidth]{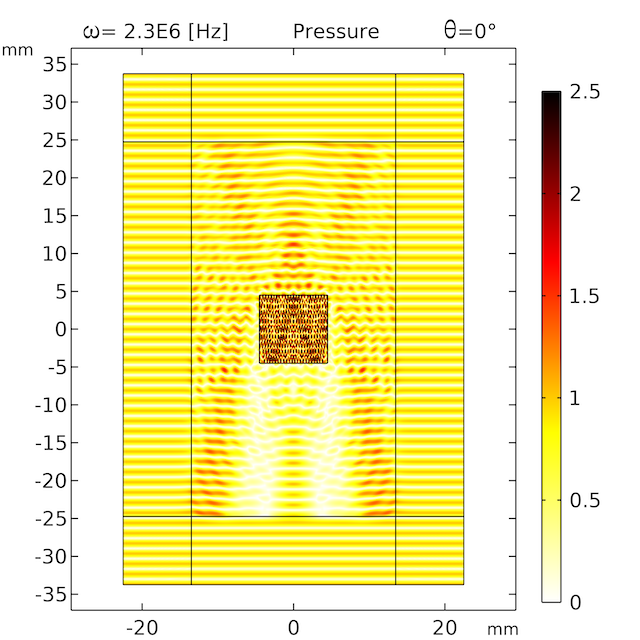}
	\end{minipage}
	\begin{minipage}[H]{0.32\textwidth}
		\includegraphics[width=\textwidth]{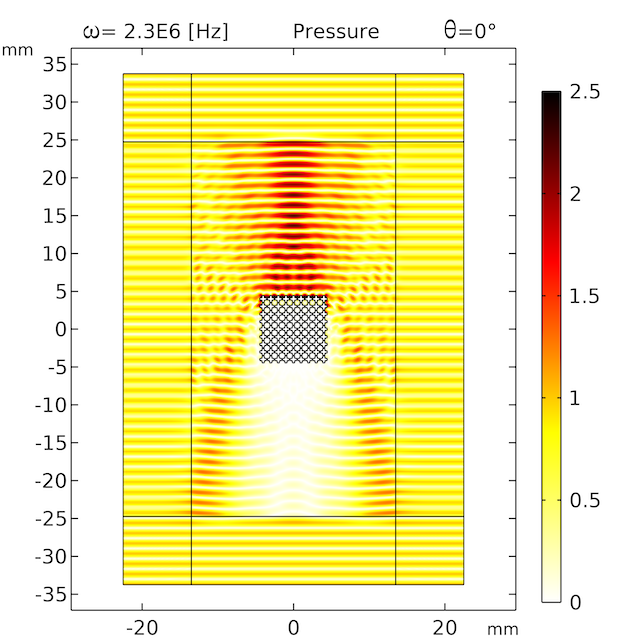}
	\end{minipage}
	\begin{minipage}[H]{0.32\textwidth}
		\includegraphics[width=\textwidth]{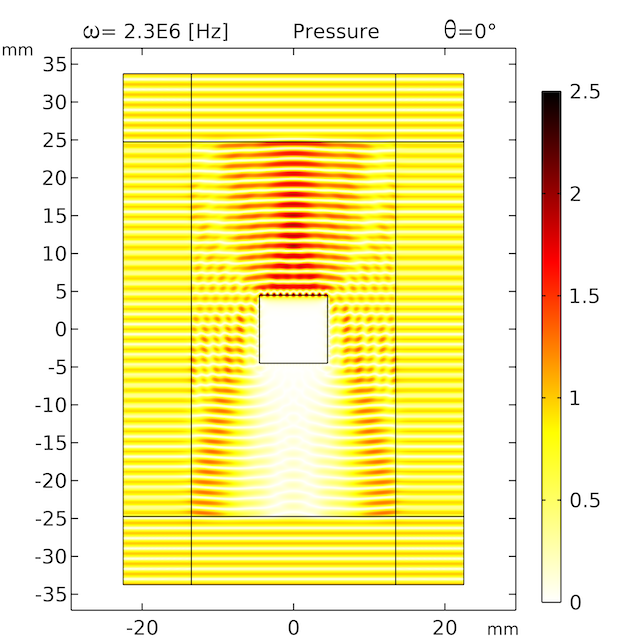}
	\end{minipage}
	\\
	\begin{minipage}[H]{0.32\textwidth}
		\includegraphics[width=\textwidth]{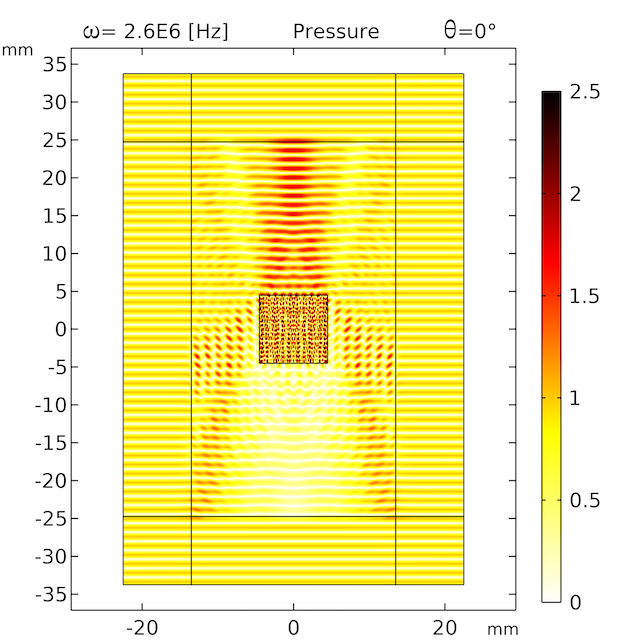}
	\end{minipage}
	\begin{minipage}[H]{0.32\textwidth}
		\includegraphics[width=\textwidth]{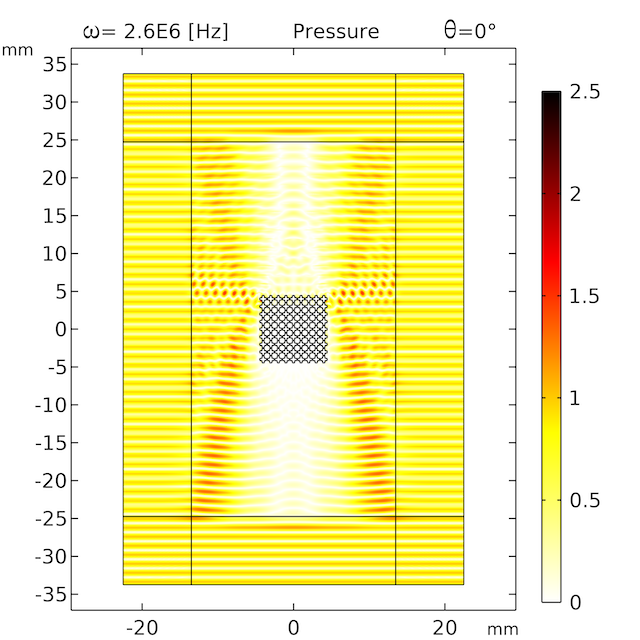}
	\end{minipage}
	\begin{minipage}[H]{0.32\textwidth}
		\includegraphics[width=\textwidth]{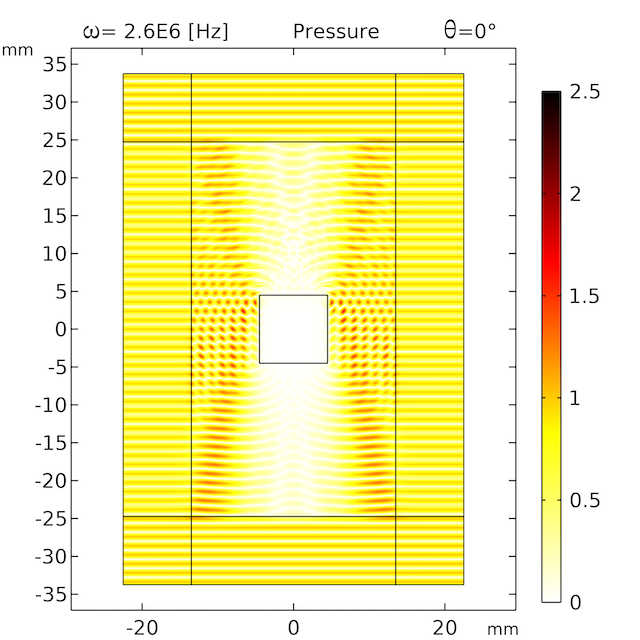}
	\end{minipage}
	\caption{
	Norm of the dimensionless displacement field (scaled with respect to the amplitude of the incident wave) for a time harmonic incident pressure wave with a 0 degrees incident angle for \textit{(first row)} a frequency of 0.83~[MHz], \textit{(second row)} a frequency of 1.2~[MHz], \textit{(third row)} a frequency of 2.3~[MHz], and \textit{(fourth row)} a frequency of 2.6~[MHz].
	In the fourth row it is possible to see how the reflected wave is in counter-phase with respect the incident one unloading the material in a cone above the square made-up of metamaterial.
	(\textit{left column}) The metamaterial is modelled with the classical Cauchy model; 
	(\textit{center column}) the metamaterial is modelled encoding all the geometrical details; 
	(\textit{right column}) the metamaterial is modelled with the relaxed micromorphic material. 
	}
	\label{fig:Cross_compa_p_a_1}
\end{figure}

We report further interesting results obtained for a normal incidence angle ($\theta=0$).
As it can be seen in the bottom images of Fig. \ref{fig:Cross_compa_p_a_1}, the reflected wave of a pressure wave with a 0 degrees incidence angle and 2.6~[MHz] frequency results in a counter-phase with respect the incident wave, resulting in an unloaded cone above the square made-up of metamaterial.
This does not happen for different frequencies as it can be seen in the previous figures.
The chosen frequencies are $\omega=\{0.83,1.2,2.3,2.6\}$~[MHz], which are below the band-gap, in the low-range, middle-range, and in the high-range of the band-gap, respectively.

\end{document}